\documentclass[%
 reprint,
superscriptaddress,
 amsmath,amssymb,
 aps,
]{revtex4-2}

\usepackage{graphicx}% Include figure files
\usepackage{dcolumn}% Align table columns on decimal point
\usepackage{bm}% bold math
\usepackage{aas_macros}
\usepackage{color}
\usepackage{amsmath}
\usepackage{natbib}

\usepackage{setspace}

\newcommand{\average}[1]{\ensuremath{\langle#1\rangle}}

\usepackage{hyperref}
\hypersetup{
    colorlinks=true,
    linkcolor=blue,
    citecolor=blue,        % color of links to bibliography
    filecolor=magenta,      
    urlcolor=blue,
}
\begin{document}

\preprint{APS/123-QED}

\title{
%The dynamics of fast neutrino flavor conversions with scattering effects
Dynamics of fast neutrino flavor conversions with scattering effects: a detailed analysis
}
% Force line breaks with \\

%\affiliation{%

\author{Hirokazu Sasaki}
\email{hsasaki@lanl.gov}
%, orcid: 0000-0001-9866-7003}
%\affiliation{Division of Science, National Astronomical Observatory of Japan, \\
%2-21-1 Osawa, Mitaka, Tokyo 181-8588, Japan}
\affiliation{Theoretical Division, Los Alamos National Laboratory, Los Alamos, New Mexico 87545, USA}
\author{Tomoya Takiwaki}
\email{takiwaki.tomoya.astro@gmail.com}
%, orcid: 0000-0003-0304-9283}
\affiliation{%
Division of Science, National Astronomical Observatory of Japan, \\
2-21-1 Osawa, Mitaka, Tokyo 181-8588, Japan}


%\affiliation{
%\affiliation{
%\affiliation{%


\date{\today}% It is always \today, today,

\begin{abstract}
 We calculate fast conversions of two flavor neutrinos by considering Boltzmann collisions of neutrino scatterings.   In an idealized angular distribution of neutrinos with electron-lepton number crossing, we find that the collision terms of the neutrino scattering enhance the transition probability of fast flavor conversions as in the previous study.  We analyze the dynamics of fast flavor conversions with collisions in detail based on the motion of polarization vectors in cylindrical coordinate analogous to a pendulum motion. The phase of the all polarization vector synchronizes in the linear evolution, and the phase deviation from the Hamiltonian governs the conversion of neutrino flavor. In the non-linear evolution, a closed orbit in the phase space is observed. The collision terms break the closed orbit and gradually make the phase space smaller.  The flavor conversions are enhanced during this limit cycle.  After the significant flavor conversion, all of the neutrino polarization vectors start to align with the $z$-axis owing to the collision effect within the time scale of the collision term irrespective of neutrino scattering angles. We also show the enhancement or suppression of the flavor conversions in various setups of the collision terms and verify consistency with previous studies.
 Though our analysis does not fully understand the dynamics of fast flavor conversion, the framework gives a new insight into this complicated phenomenon in further study.

%We calculate fast conversions of two flavor neutrinos by considering Boltzmann collisions of neutrino scatterings.
%In an idealized angular distribution of neutrino with electron-lepton number crossing,
%the collision terms of the neutrino scattering enhance the transition probability of fast flavor conversions as in the previous study.
%We analyze the dynamics of fast flavor conversions with neutrino collisions in detail based on the motion of neutrino polarization vectors analogous to a pendulum motion in cylindrical coordinate.
%The collision terms break a closed orbit and show a limit cycle in the motion of neutrino polarization vectors.
%The flavor conversions are enhanced during the limit cycle.
%After the significant conversion, all of neutrino polarization vectors start to align with $z$-axis owing to the collision effect within the time scale of the collision term.
%We find that the values of $z$-components of neutrino polarization vectors are equivalent irrespective of neutrino scattering angles.
%Though our analysis does not fully understand the dynamics of fast flavor conversion, we hope the analysis framework gives a new insight into this phenomenon.
%Fast flavor conversions finally settle down into equilibrium when fluxes of all species of neutrinos become isotropic.
\end{abstract}

\maketitle

%\tableofcontents

%%%%%%%%%%%%%%%%%%%%%%%%%%%%%%%%%%%%%%%%
\section{Introduction}
%%%%%%%%%%%%%%%%%%%%%%%%%%%%%%%%%%%%%%%%
 
There are many sources of neutrinos in our nature (see, e.g., a review in Ref.~\cite{Vitagliano:2019yzm}). The detection of neutrinos from explosive astrophysical sites such as core-collapse supernovae (CCSNe), and neutron star mergers helps understand the mechanism of their explosive phenomena. The neutrino fluxes are affected by neutrino oscillations that are sensitive to environment inside the astrophysical sites (see, e.g., reviews in Refs.~\cite{Horiuchi2018WhatDetection,Janka2017NeutrinoSupernovae,Mirizzi2015SupernovaDetection}). Coherent forward scatterings of neutrinos with background matter induce a refractive effect that has an influence on flavor conversions of neutrinos. The matter effect called the ``Mikheyev-Smirnov-Wolfenstein (MSW) effect" \cite{Wolfenstein1978NeutrinoMatter,Mikheev:1986gs} is caused by charged current interactions of neutrinos with background electrons. Neutrino-neutrino interactions play important role in flavor conversions in dense neutrino gas , where large numbers of neutrinos are produced. It is numerically found that neutrino-neutrino interactions induce non-linear flavor conversions so-called ``Collective neutrino oscillations" (CNOs) in CCSNe \cite{Duan:2006an,Fogli:2007bk,Dasgupta:2010cd,Duan:2010bf,Friedland:2010sc,Mirizzi:2010uz,Cherry:2010yc,Chakraborty:2011gd,Wu:2014kaa,Cirigliano:2017hmk,Cirigliano:2018rst,Sasaki:2017jry,Sasaki:2019jny,Cherry:2019vkv,Zaizen2020JCAP,Zaizen:2020xum}, neutron star mergers \cite{Malkus:2014iqa,Malkus:2015mda,Wu:2015fga,Zhu:2016mwa,Frensel:2016fge,Chatelain:2016xva,Tian:2017xbr,Vlasenko:2018irq,Shalgar:2017pzd}, and the early Universe \cite{Kostelecky:1993yt,Dolgov:2002ab,Wong:2002fa,Johns:2016enc,deSalas:2016ztq,Hasegawa:2019jsa,Hansen:2020vgm}. The collective neutrino oscillations can change neutrino spectra dramatically and potentially affect neutrino signals in neutrino detectors \cite{Wu:2014kaa,Sasaki:2019jny,Zaizen2020JCAP,Zaizen:2020xum} and nucleosynthesis inside astrophysical sites such as the $\nu p$ process \cite{MartinezPinedo:2011br,Pllumbi:2014saa,Sasaki:2017jry,Xiong:2020ntn} and the $r$ process \cite{Malkus:2015mda,Wu:2017drk,George:2020veu}.

Recently, much attention has been focused on fast-pair wise collective neutrino oscillations whose oscillation scales are
$\sim\mathcal{O}(10^{-5})\,\mathrm{km}$ (see e.g. a review \cite{Tamborra:2020cul}). The fast flavor conversions may occur near neutrino spheres and have strong impact on explosion dynamics in astrophysical sites. The possibility of fast flavor conversions was first proposed in Ref.~\cite{Sawyer:2005jk}. Fast flavor conversions are associated with anisotropic angular distribution of neutrinos \cite{Sawyer:2005jk,Sawyer:2008zs,Sawyer:2015dsa}. Such fast modes are not confirmed in numerical simulations assuming the ``bulb model" \cite{Duan:2006an} where all of neutrinos are emitted isotropically on the surface of a fixed neutrino sphere irrespective of neutrino species and neutrino energy. The growth of instability induced by fast flavor conversions is studied through the linear stability analysis \cite{Dasgupta:2016dbv,Izaguirre:2016gsx,Chakraborty:2016lct,Capozzi:2017gqd}. The instability of fast flavor conversions is studied in CCSNe \cite{Abbar:2018shq,Abbar:2019zoq,Glas:2019ijo,DelfanAzari:2019tez,DelfanAzari:2019epo,Morinaga:2019wsv,Nagakura2019a,Abbar:2020qpi} and  neutron star mergers \cite{Wu:2017qpc,George:2020veu,Li:2021vqj} by employing simulation data of neutrino radiation hydrodynamics. The crossing between angular distribution of $\nu_{e}$ and that of $\bar{\nu}_{e}$ so-called ``electron-lepton number (ELN) crossing" is associated with fast flavor conversions \cite{Abbar:2017pkh,Izaguirre:2016gsx}. The methods of ELN-crossings search in multi-dimensional (multi-D) CCSNe simulations are recently developed \cite{Abbar:2020fcl,Nagakura:2021suv,Johns:2021taz} and such treatments are applied to state-of-the-art supernova simulations \cite{Capozzi:2020syn,Nagakura2021a}. In general, numerical simulations of fast flavor conversions in non-linear regime are challenging problem due to the numerical difficulties, but the fast flavor conversions can be calculated even in non-linear regime within local simulations
\cite{Dasgupta:2017oko,Abbar:2018beu,Capozzi:2018clo,Johns:2019izj,Shalgar:2020xns,Shalgar:2020wcx,Shalgar:2021wlj,Martin:2019gxb,Martin:2021xyl,Zaizen2021a,Xiong:2021dex,Kato2021a,Wu2021a,Richers:2021nbx,Johns:2021arXiv210411369J,Shalgar:2021arXiv210615622S}.

The Boltzmann collision terms that correspond to contributions from incoherent scatterings, emission, and absorption of neutrinos are taken into account in simulations of collective neutrino oscillations \cite{Cirigliano:2017hmk,Cirigliano:2018rst,Zaizen2020JCAP,Capozzi:2018clo,Shalgar:2020wcx,Shalgar:2021wlj,Martin:2021xyl,Kato2021a}.Neutrino scattering terms can increase the transition probability of neutrinos after the fast flavor conversions \cite{Shalgar:2020wcx,Johns:2021arXiv210411369J}. On the other hand, Ref.~\cite{Martin:2021xyl} shows that fast flavor conversions are damped and neutrino spectra become isotropic on the scale of the mean free path of neutrinos. These two results seem to contradict with each other and the role of Boltzamann collision on fast flavor conversions is still unknown.

%Neutrino scattering terms can increase the transition probability of neutrinos and neutrino spectra become anisotropic after the fast flavor conversions \cite{Shalgar:2020wcx}. On the other hand, Ref.~\cite{Martin:2021xyl} shows that fast flavor conversions are damped and neutrino spectra become isotropic on the scale of the mean free path of neutrinos. These two results seem to contradict with each other and the role of Boltzamann collision on fast flavor conversions is still unknown.

In this work, we calculate fast flavor conversions in the non-linear regime and study the effect of neutrino scatterings on the fast flavor conversions based on the dynamics of two flavor neutrino polarization vectors. In Sec.~\ref{sec:Methods}, we explain our numerical setup for fast flavor conversions in the non-linear regime. In Sec.~\ref{sec:Results}, we show the numerical results and discuss the effect of neutrino scatterings. Here, we divide the evolution of fast flavor conversions into three stages: linear evolution phase, limit cycle phase, and relaxation phase. We also discuss effects of various collision terms on fast flavor conversions. We finally summarize our result in Sec.~\ref{sec:Conclusions}.

%%%%%%%%%%%%%%%%%%%%%%%%%%%%%%%%%%%%%%%%
\section{Methods}\label{sec:Methods}
%%%%%%%%%%%%%%%%%%%%%%%%%%%%%%%%%%%%%%%%

We calculate fast flavor conversions of two flavor neutrinos ($\nu_{e}$,$\nu_{x}$) and antineutrinos ($\bar{\nu}_{e}$,$\bar{\nu}_{x}$) with collision terms of neutrino scattering based on a formalism in Ref.~\cite{Shalgar:2020wcx}. We analyze behaviors of the flavor conversions through the geometrical representation of neutrino density matrices. In this section, we explain numerical setup for our calculation and introduce equation of motion of polarization vectors to analyze the dynamics of flavor conversions.

 Neutrino oscillations considering Boltzmann collisions are expressed by the time evolution of neutrino density matrix $\rho$ and that of antineutrino $\bar{\rho}$ \cite{Sigl:1992fn,Yamada:2000za,Cirigliano:2009yt,Vlasenko:2013fja,Blaschke:2016xxt,Cirigliano:2018rst,Shalgar:2020wcx,Martin:2021xyl}:
\begin{align}
     \frac{\mathrm{d}}{\mathrm{d} t}\rho=&-i[H,\rho] + C[\rho,\bar{\rho}],\\
    \frac{\mathrm{d}}{\mathrm{d} t}\bar{\rho}=&-i[\bar{H},\bar{\rho}] + \bar{C}[\rho,\bar{\rho}],
\end{align}
where $H(\bar{H})$ and $C(\bar{C})$ are Hamiltonian and collision term of neutrinos (antineutrinos), respectively. The time integration of the equations is performed by Runge-Kutta 4th method. 
We have confirmed that the results with the 5th-order method are the same as that of the 4th-order method.
The time step is chosen as $\Delta t < 0.1/\max[H_{i,j,\theta},\bar{H}_{i,j,\theta}]$, where $i,j$ are $e,x$; and the components of Hamiltonian depend on the angle, $\theta$.

In explosive astrophysical sites without magnetic fields, neutrino Hamiltonian is composed of vacuum Hamiltonian, the MSW matter term and potential of neutrino-neutrino interactions \cite{Sasaki:2021bvu}. For simplicity, the MSW matter potential is ignored in our calculation. Instead of the matter potential, we impose an effective vacuum mixing angle: $\theta_{\mathrm{v}}=10^{-6}$ that compensates the effect of matter suppression \cite{Mirizzi:2010uz}. The vacuum Hamiltonian of two flavor neutrino is described by
\begin{equation}
    H_{\mathrm{vac}}=\frac{\omega}{2}\left(
\begin{array}{c c}
-\cos2\theta_{\mathrm{v}}&\sin2\theta_{\mathrm{v}}\\
  \sin2\theta_{\mathrm{v}}&\cos2\theta_{\mathrm{v}}
\end{array}
\right),
\end{equation}
where $\omega=\frac{\Delta m^{2}}{2E}$ is a vacuum frequency composed of a neutrino energy $E$ and neutrino mass difference $\Delta m^{2}$. We use the small mass difference: $\Delta m^{2}=2.5\times10^{-6}\,\mathrm{eV}^{2}$ which leads to a periodic trend of fast flavor conversions \cite{Dasgupta:2017oko,Johns:2019izj,Shalgar:2020xns}. The fast flavor conversions are associate with angular dependence in neutrino distributions.
In this work, We remove energy dependence in neutrino distribution and focus on flavor conversions of single energy neutrinos ($E=50$\,MeV).
In the case of azimuthal symmetric neutrino emission, the potential of neutrino-neutrino interaction is given by the integration over a polar angle $\theta$ \cite{Shalgar:2020xns},
\begin{align}
    H_{\nu\nu}(\cos\theta)=2\pi\mu&\int^{1}_{-1}\mathrm{d}\cos\theta^{\prime}\left(
    1-\cos\theta^{\prime}\cos\theta
    \right)\nonumber\\
    &\left\{
    \rho(\cos\theta^\prime)-\bar{\rho}(\cos\theta^\prime)
    \right\},
\end{align}
where $\mu$ is the strength of neutrino-neutrino interactions. Throughout the calculation, we fix the strength of neutrino-neutrino interaction as $\mu=10^{4}\,\mathrm{km}^{-1}$. 
To minimize the error in the integration, we employ 2000 Gauss-Legendre mesh for $\theta$.

In general, the Boltzmann collision of neutrino scatterings should depend on neutrino energy, scattering angles and flavors. In this work, we ignore the flavor dependence in collision terms and focus on elastic scattering of neutrinos. We employ direction-changing collisions \cite{Shalgar:2020wcx},
\begin{align}
    C[\rho,\bar{\rho}]=&-\int^{1}_{-1}\mathrm{d}\cos\theta^{\prime}C_{\mathrm{loss}}\rho(\cos\theta)\nonumber\\
    &+\int^{1}_{-1}\mathrm{d}\cos\theta^{\prime}C_{\mathrm{gain}}\rho(\cos\theta^{\prime}), \label{eq:collision rho}\\
    \bar{C}[\rho,\bar{\rho}]=&-\int^{1}_{-1}\mathrm{d}\cos\theta^{\prime}\bar{C}_{\mathrm{loss}}\bar{\rho}(\cos\theta)\nonumber\\
    &+\int^{1}_{-1}\mathrm{d}\cos\theta^{\prime}\bar{C}_{\mathrm{gain}}\bar{\rho}(\cos\theta^{\prime}),\label{eq:collision rhob}
\end{align}
where the first terms and second terms of the above equations represent ``loss" and ``gain" terms, respectively. The number of neutrinos (antineutrinos) are conserved when $C_{\mathrm{loss}}=C_{\mathrm{gain}}$ ($\bar{C}_{\mathrm{loss}}=\bar{C}_{\mathrm{gain}}$) is satisfied. Here, we assume constant collision terms: $C_{\mathrm{loss}}=C_{\mathrm{gain}}=\bar{C}_{\mathrm{loss}}=\bar{C}_{\mathrm{gain}}=C/2$ irrespective of neutrino scattering angles as followed in Refs.~\cite{Shalgar:2020wcx,Shalgar:2021wlj}.

In the beginning of the calculation ($t=0$), we employ the distributions of $\nu_{e}$ and $\bar{\nu}_{e}$ such as
\begin{equation}
\label{eq:initial condition}
\begin{split}
    &\rho_{ee}(\cos\theta)=0.5,\\
    &\bar{\rho}_{ee}(\cos\theta)=0.47+0.05\exp(-2(\cos\theta-1)^{2}),
\end{split}
\end{equation}
where $\rho_{xx}$ and  $\bar{\rho}_{xx}$ are equal to zero (see the top panel of Fig.~\ref{fig:spec relaxation_caseb}). There is a ELN crossing around $\cos\theta\sim0.5$ in Eq.~(\ref{eq:initial condition}). These initial conditions correspond to those in the Case B in Ref.~\cite{Shalgar:2020wcx}. We impose a random phase, i,e., $\rho_{ex}=\rho_{ee} \epsilon$, $\bar{\rho}_{ex}=\bar{\rho}_{ee} \epsilon$,where $\epsilon$ is a random complex number of the order of $10^{-8}$.

In our numerical setup, the neutrino density matrix depends on the polar scattering angle $\theta$ and the time $t$.
Hereafter, for the simple notation, we do not write the dependence explicitly. In two flavor neutrinos, the neutrino density matrix is decomposed of Pauli matrices $\sigma_{i}(i=x,y,z)$ and polarization vector $\mathbf{P}=(P_{x},P_{y},P_{z})$: 
\begin{equation}
\label{eq:geometrical representation}
\rho=\left(
\begin{array}{c c}
\rho_{ee}&\rho_{ex}\\
\rho_{xe}&\rho_{xx}
\end{array}
\right)
=\frac{\mathrm{Tr}\rho}{2}I_{2\times2}+\frac{P_{i}\sigma_{i}}{2},
\end{equation}
where $I_{2\times2}=\mathrm{diag}(1,1)$. The density matrix of antineutrino $\bar{\rho}$ is also represented by the polarization vector of antineutrino $\mathbf{\bar{P}}=(\bar{P}_{x},\bar{P}_{y},\bar{P}_{z})$ in the same way. In our numerical setup, the equations of motion of polarization vectors are written as
%and that of $\mathrm{Tr}\rho$ are written as
%\begin{equation}
%\frac{\mathrm{d}}{\mathrm{d}t}\mathrm{Tr}\rho=-C\left(
%\mathrm{Tr}\rho-\average{\mathrm{Tr}\rho}
%\right)
%\end{equation}
\begin{equation}
\begin{split}
\label{eq:EOM neutrino vector}
\frac{\mathrm{d}}{\mathrm{d}t}\mathbf{P}&=\left(
    +\omega\mathbf{B}+\mu^{\prime}\mathbf{D_{0}}-\mu^{\prime}\cos\theta\mathbf{D_{1}}
    \right)\times\mathbf{P}-C\mathbf{P}+C\average{\mathbf{P}},\\
\frac{\mathrm{d}}{\mathrm{d}t}\mathbf{\bar{P}}&=\left(
    -\omega\mathbf{B}+\mu^{\prime}\mathbf{D_{0}}-\mu^{\prime}\cos\theta\mathbf{D_{1}}
    \right)\times\mathbf{\bar{P}}-C\mathbf{\bar{P}}+C\average{\mathbf{\bar{P}}}.
\end{split}
\end{equation}
The variables in the equations are defined as
\begin{equation}
\begin{split}
\label{eq:vectors}
    \mathbf{B}&=\left(
    \sin2\theta_\mathrm{v},0,-\cos2\theta_\mathrm{v}
    \right),\\
    \mathbf{D_{0}}&=\average{\mathbf{P}-\mathbf{\bar{P}}},\\
    \mathbf{D_{1}}&=\average{\left(
    \mathbf{P}-\mathbf{\bar{P}}
    \right)\cos\theta},
\end{split}
\end{equation}
where $\mu^{\prime}=4\pi\mu$. Here, $\average{A}=\frac{1}{2}\int^{1}_{-1} A\ \mathrm{d}\cos\theta$ represents the angular average of a quantity $A$ that is a function of $\cos\theta$. In the next Section, we analyze the behaviors of neutrino oscillations based on the motion of the polarization vectors governed by Eq.~(\ref{eq:EOM neutrino vector}). The $z$-components of polarization vectors include information of numbers of neutrinos. The finite value of $C$ changes the length of the neutrino polarization vector. From Eq.~(\ref{eq:EOM neutrino vector}), we can derive the equation of motion of the angled averaged length of polarization vectors,
\begin{equation}
\begin{split}
\label{eq:mean square polarization vectors}
\frac{\mathrm{d}}{\mathrm{d}t}\average{|P|^{2}}&=-2C\left(
\average{|P|^{2}}-|\average{P}|^{2}
\right),\\
\frac{\mathrm{d}}{\mathrm{d}t}\average{|\bar{P}|^{2}}&=-2C\left(
\average{|\bar{P}|^{2}}-|\average{\bar{P}}|^{2}
\right).
\end{split}
\end{equation}
The right hand sides of Eq.~\eqref{eq:mean square polarization vectors} vanish when the deviations in the angular distributions of neutrino polarization vector and that of antineutrinos disappear. The Eq.~(\ref{eq:mean square polarization vectors}) implies that the distributions of polarization vectors become isotropic in equilibrium states of fast flavor conversions owing to collision effects. Furthermore, the Boltzmann collision changes the direction of emitted neutrinos, so that, in general, the $\mathrm{Tr}\rho$ and $\mathrm{Tr}\bar{\rho}$ are no longer invariant during flavor conversions even though the total neutrino numbers $2\average{\mathrm{Tr}\rho}$ and $2\average{\mathrm{Tr}\bar{\rho}}$ are conserved. The time evolution of these traces of density matrices can be solved analytically. In the case of initial distribution as shown in  Eq.~(\ref{eq:initial condition}), the trace of neutrino density matrix does not evolve ($\mathrm{Tr}\rho=0.5$). On the other hand, the value of $\mathrm{Tr}\bar{\rho}$ exponentially approach to that of $\average{\mathrm{Tr}\bar{\rho}}(=0.4857)$ irrespective of scattering angle $\theta$ and the angular dependence finally disappears in equilibrium.

%%%%%%%%%%%%%%%%%%%%%%%%%%%%%%%%%%%%%%%%
\section{Results}\label{sec:Results}
%%%%%%%%%%%%%%%%%%%%%%%%%%%%%%%%%%%%%%%%

\begin{figure}[t]
\includegraphics[width=0.95\linewidth]{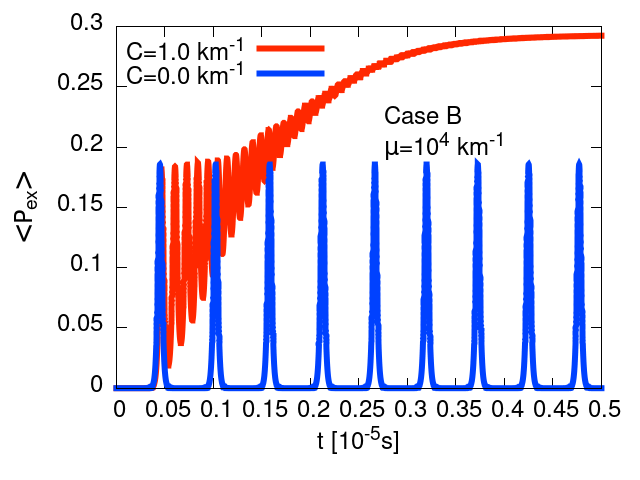}\\
\caption{%%
The time evolution of transition probability $\average{P_{ex}}$ with (red) and without (blue) the collision terms.
}
\label{fig:caseB-t-Pex}
\end{figure}

Figure \ref{fig:caseB-t-Pex} shows the overall evolution of angle averaged transition probability $\average{P_{ex}}$ \cite{Shalgar:2020wcx},
\begin{equation}
\label{eq:transition probability general}
    \average{P_{ex}}=\frac{\average{\rho_{ee}}_{\rm ini}-\average{\rho_{ee}}}{\average{\rho_{ee}}_{\rm ini}-\average{\rho_{xx}}_{\rm ini}},
\end{equation}
where $\average{\rho_{\alpha\alpha}}_{\rm ini}$ is the initial value of $\average{\rho_{\alpha\alpha}}(\alpha=e,x)$. In our numerical setup with Eqs.~(\ref{eq:collision rho}) and (\ref{eq:collision rhob}), the trace of neutrino density matrix is invariant during flavor conversions ($\mathrm{Tr}\rho=0.5$), so that the $\average{P_{ex}}$ is described by
\begin{equation}
\label{eq:transition probability}
    \average{P_{ex}}=\frac{1}{2}\left(
    1-\frac{\average{P_{z}}}{\average{P_{z}}_{\rm ini}}
    \right),
\end{equation}
where $\average{P_{z}}_{\rm ini}$ is the initial value of $\average{P_{z}}$, $0.5$.
%From Eq.~(\ref{eq:geometrical representation}) and time invariant trace of neutrino density matrix ($\mathrm{Tr}\rho=0.5$), 
In the case of $C=0\,\mathrm{km}^{-1}$ (blue curve), the periodic structure of fast flavor conversions is confirmed. On the other hand, in the case of $C=1\,\mathrm{km}^{-1}$ (red curve), flavor conversions are enhanced by the collision terms and the transition probability reaches an equilibrium value. Properties of fast flavor conversions in our calculation are consistent with results in Ref.~\cite{Shalgar:2020wcx} (see the Case B of Fig.~$2$ in the paper) except for the time scale of flavor conversions. The probability in Fig.~\ref{fig:caseB-t-Pex} becomes large when $t>0.05\times10^{-5}\,{\rm s}$ in both cases. The flavor conversions in Fig.~\ref{fig:caseB-t-Pex} evolve faster than those of calculations in Ref.~\cite{Shalgar:2020wcx}. Such discrepancy in the time scale of flavor conversions is also reported in QKE-MC simulations \cite{Kato2021a}. This issue might be related to the strength of the initial perturbation. Since we imposed $\sim 10^{-8}$ random seed, the first peak in $\average{P_{ex}}$ may arise faster.
%with high time resolution $\Delta t$

The dynamics of fast flavor conversions is mainly divided by three epochs such as the linear evolution phase, the limit cycle phase, and the relaxation phase. The fast flavor conversions are enhanced in the limit cycle phase owing to the collision effect. The distribution of neutrinos finally become isotropic in the relaxation phase when the evolution time is comparable with the time scale of the collision term ($c t\sim C^{-1}$), where $c$ is the speed of light.
Hereafter we omit $c$ when we  convert timescale to length scale.

\subsection{Linear evolution phase}

\begin{figure}[t]
\includegraphics[width=0.95\linewidth]{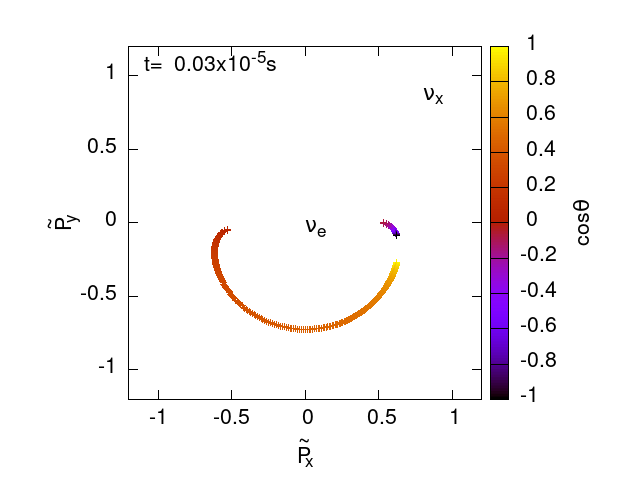}\\
\caption{%%
The map of normalized neutrino polarization vector $(\Tilde{P}_{x},\Tilde{P}_{y})$ in the linear evolution phase ($t=0.03\times10^{-5}$\,s). The normalized polarization vector is defined in Eq.~(\ref{eq:polarization vector normalized}).
}
\label{fig:palog00905_caseb}
\end{figure}

We call the epoch before reaching the first peak in $\average{P_{ex}}$ linear evolution phase. In this early phase of fast flavor conversions ($t<0.05\times10^{-5}$\,s), the instability of flavor conversions appears near the ELN crossing in the initial distributions of $\nu_{e}$ and $\bar{\nu}_{e}$ ($\cos\theta\sim0.5$). As shown in Fig.~\ref{fig:caseB-t-Pex}, collision effects are negligible in the linear evolution phase because of the large time scale of the collision terms ($t \ll C^{-1}$), so that flavor conversions without and with collision terms are almost equivalent. Here, we use time snapshots of neutrino polarization vectors with $C=1\,\mathrm{km}^{-1}$ at $t=0.03\times10^{-5}$\,s to study behaviors of fast flavor conversions during the linear phase.

Fig.~\ref{fig:palog00905_caseb} shows a map of normalized neutrino polarization vectors on the $\Tilde{P}_{x}-\Tilde{P}_{y}$ plane at $t=0.03\times 10^{-5}$\,s. In this figure, for the convenience to see a polarization vector in the case of a small transition probability, the polarization vector is normalized as
\begin{equation}
\begin{split}
\label{eq:polarization vector normalized}
    \Tilde{P}_{x}&=\left(1+\frac{\log_{10}P_{R}/|P|}{15}\right)
    \cos (P_\phi),\\
    \Tilde{P}_{y}&=\left(1+\frac{\log_{10}P_{R}/|P|}{15}\right)
    \sin (P_\phi),
    \end{split}
\end{equation}
where $P_{R}=\sqrt{P_{x}^{2}+P_{y}^{2}}$ and $P_\phi=\tan^{-1}(P_{y}/P_{x})$ (in this paper, inverse tangent is calculated by \verb!atan2! function in \verb!C++!). At first, all neutrino polarization vectors lie in the $z-$axis ($(0,0)$ in Fig.~\ref{fig:palog00905_caseb}). As the calculation time has passed, the polarization vector begins a spiral motion around the $z-$axis increasing the value of $P_{R}$. 

\begin{figure}[htb]
\includegraphics[width=0.85\linewidth]{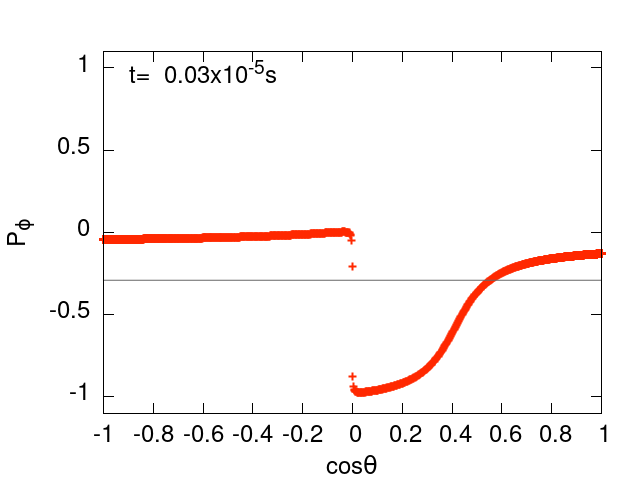}\\
\includegraphics[width=0.85\linewidth]{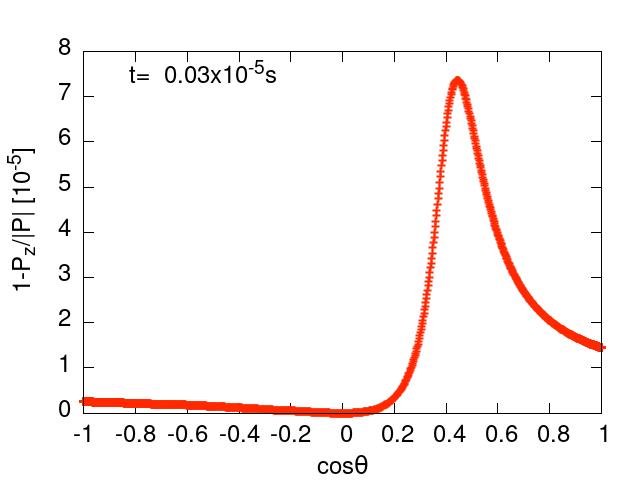}\\
\includegraphics[width=0.85\linewidth]{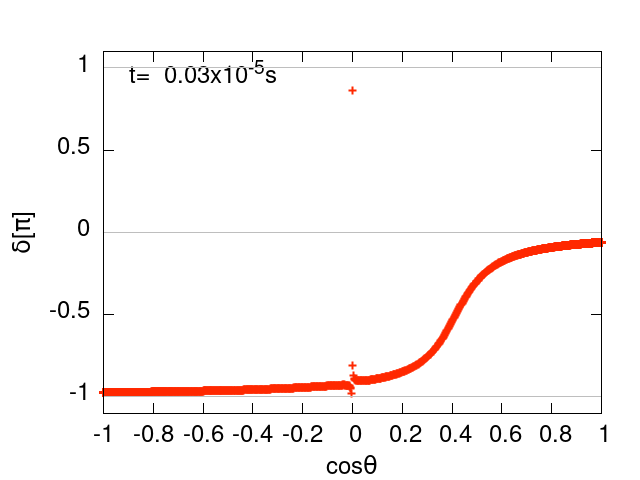}
\caption{%%
Polarization vector at $t=0.03\times10^{-5}$\,s.
Top: The angular distribution of the phase $P_\phi=\tan^{-1}(P_{y}/P_{x})$.
The gray line is the phase of averaged polarization vector, $\langle P\rangle_\phi=\tan^{-1}(\langle P_y\rangle/\langle P_x\rangle)$.
Middle: The angular distribution of $1-P_{z}/|P|$ in the unit of $10^{-5}$.
Bottom: The angular distribution of the phase difference $\delta$. The definition of the $\delta$ is given in Eq.~(\ref{eq:phase difference}).
The gray lines indicate $1,0,-1$\,$[\pi]$.
}
\label{fig:phase00905_caseb}
\end{figure}

The dynamics of the polarization can be understood in the cylindrical coordinate ($R$, $\phi$, $z$) as follows. First, we need to define and explain some variables.
The quantity $P_\phi$ represents a phase of polarization vector on the $\Tilde{P}_{x}$-$\Tilde{P}_{y}$ plane. In negative value of $\cos\theta$, the motions of polarization vectors synchronize with each other, which results in the same value of $P_\phi$ in the top panel of Fig.~\ref{fig:phase00905_caseb}. On the other hand, the phase $P_\phi$ is distributed broadly in the case of $\cos\theta>0$. The phase of $\cos\theta>0.6$ is close to that of $\cos\theta<0$. The $P_\phi$ becomes opposite phase in $0<\cos\theta<0.2$. The value of $P_{R}$ becomes small around $\cos\theta\sim0$, so that $\theta=\pi/2$ can be regarded as a singular point for $P_\phi$. We discuss the reason why the $P_{R}$ is small around $\cos\theta=0$ in Sec.\ref{sec:Relaxation phase}.
The $z$-component of the neutrino polarization vector is related to the transition probability. According to Eq.~(\ref{eq:transition probability}), the transition probability increases as the value of $P_{z}$ decreases. The middle panel of Fig.~\ref{fig:phase00905_caseb} shows the distribution of $1-P_{z}/|P|$ at $t=0.03\times10^{-5}$\,s. The flavor conversions do not proceed in $\cos\theta<0.2$. However,
the flavor conversions become prominent in $\cos\theta\sim0.4$ that is close to the angle of the ELN crossing ($\cos\theta\sim 0.5$) in the initial distributions.
Here, we define the phase difference between $\mathbf{P}$ and the polarization vector of neutrino Hamiltonian $\mathbf{H}$ on the $x$-$y$ plane,
\begin{equation}
\label{eq:phase difference}
\begin{split}
    H_{R}&=\sqrt{H_{x}^{2}+H_{y}^{2}},\\
    \delta&=P_\phi-H_\phi,
    \end{split}
\end{equation}
where $\mathbf{H}=\omega\mathbf{B}+\mu^{\prime}\mathbf{D_{0}}-\mu^{\prime}\cos\theta\mathbf{D_{1}}$ is the polarization vector of neutrino Hamiltonian and $H_\phi=\tan^{-1}(H_{y}/H_{x})$ is the phase of $\mathbf{H}$ on the $x$-$y$ plane.The bottom panel of Fig.~\ref{fig:phase00905_caseb} shows the phase difference $\delta$ in Eq.~(\ref{eq:phase difference}) at $t=0.03\times10^{-5}$ s. The neutrino polarization vectors are almost antiparallel to the Hamiltonian vector in $\cos\theta<0.2$ because of $\cos\delta\sim-1$. The polarization vector of the neutrino Hamiltonian on the $x$-$y$ plane can be written as Eqs.~\eqref{eq:Hamiltonian vector x}~and~\eqref{eq:Hamiltonian vector y}. Since $H_x$ and $H_y$ are proportional to $\cos\theta$, $|H_{\phi}(\cos\theta\to+0)-H_{\phi}(\cos\theta\to-0)|\sim\pi$ is satisfied.
Therefore, from Eq.~\eqref{eq:phase difference}, the jump ,$|P_{\phi}(\cos\theta\to+0)-P_{\phi}(\cos\theta\to-0)|\sim\pi$ appears around $\cos\theta=0$ as shown in the top panel of Fig.~\ref{fig:phase00905_caseb}.
%The jump of $\delta$ around $\cos\theta=0$ reflects that of $P_{\phi}$ in the top panel of Fig.~\ref{fig:phase00905_caseb}.

The evolution of $P_{z}$ is related to the phase difference $\delta$. In the linear evolution phase, the contribution of the collision term is negligible, so that the time evolution of $P_{z}$ is derived from Eq.~(\ref{eq:EOM neutrino vector}),
\begin{equation}
\label{eq:time evolution Pz in the linear phase}
    \frac{\mathrm{d}}{\mathrm{d}t}P_{z} \sim H_{x}P_{y}-H_{y}P_{x}=H_{R}P_{R}\sin\delta.
\end{equation}
The value of $\sin\delta$ in Eq.~(\ref{eq:time evolution Pz in the linear phase}) is negligible when the direction of $\mathbf{P_{R}}=(P_{x},P_{y})$ is parallel ($\delta=0$) or antiparallel ($\delta=\pm\pi$) to that of $\mathbf{H_{R}}=(H_{x},H_{y})$ on the $x$-$y$ plane. At $t=0.03\times10^{-5}$\,s, the value of $|\sin\delta|$ is no longer negligible in $\cos\theta>0.2$ and becomes maximum in $\cos\theta\sim0.4$ (see the bottom panel of Fig.~\ref{fig:phase00905_caseb}), so that fast flavor conversions proceed prominently in $\cos\theta\sim0.4$ as shown in the middle panel of Fig.~\ref{fig:phase00905_caseb}.

Several studies connect the dynamics of neutrino oscillation with the synchronization phenomena \cite{Pantaleone:1998xi,Raffelt:2010za,Akhmedov:2016gzx}.
The evolution of the phase of polarization vector can be interpreted in the framework of the Kuramoto model \cite{Kuramoto2012}, i.e.,
\begin{align}
    \frac{{\rm d} \phi_i }{{\rm d} t} = \omega_i +\frac{K}{N}\sum_{j=1}^N \sin (\phi_j-\phi_i),
\end{align}
where $\phi_i$ is the phase of $i$-th oscillator, which rotates with the frequency, $\omega_i$.
The total number of oscillators is $N$ and $K$ is the coupling constant.
Sufficiently high $K$ makes synchronization, i.e., all $\phi_i$ rotates with the same frequency irrespective to the original $\omega_i$.
In the context of neutrino oscillation, this is a trivial equilibrium and no flavor conversions happen.
The flavor conversion is expected when K becomes slightly lower, and the perfect synchronization is broken \cite{Raffelt:2010za}. Note that, Refs.~\cite{Abrams2004,Kuramoto2002} consider a more complicated functions, which resemble our setup.

In our case, the strong coupling constant, $\mu$, synchronize the phase: the polarization vector at $\cos\theta>0.6$ and $\cos\theta<0$ rotates with the same phase (see Fig.~\ref{fig:phase00905_caseb} top panel).
Due to the angular distribution with ELN crossing, that at $0<\cos\theta<0.2$ rotates with a different phase but the shape of the $\delta$ is kept in the linear phase, and this part is also synchronized to the average frequency.
As a result of the synchronization, $\delta$ is kept to $-\pi/2$ at $\cos\theta \sim 0.4$, and it plays an essential role in the flavor conversion.
The profile of $\delta$ in Fig.~\ref{fig:phase00905_caseb} is clearly explained by the condition of the synchronization, see Appendix~\ref{sec:slp 1} for the details.

\begin{figure}[htbp]
\includegraphics[width=0.95\linewidth]{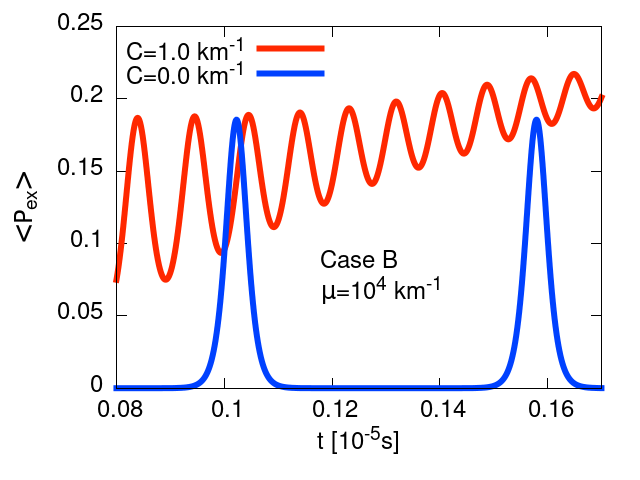}\\
\caption{%%
Enlarge view of Fig.~\ref{fig:caseB-t-Pex} in the limit cycle phase.
}
\label{fig:t-Pex_zoom_casseb}
\end{figure}

\subsection{Limit cycle phase}\label{sec:Limit cycle phase}

\begin{figure}[htbp]
\includegraphics[width=0.95\linewidth]{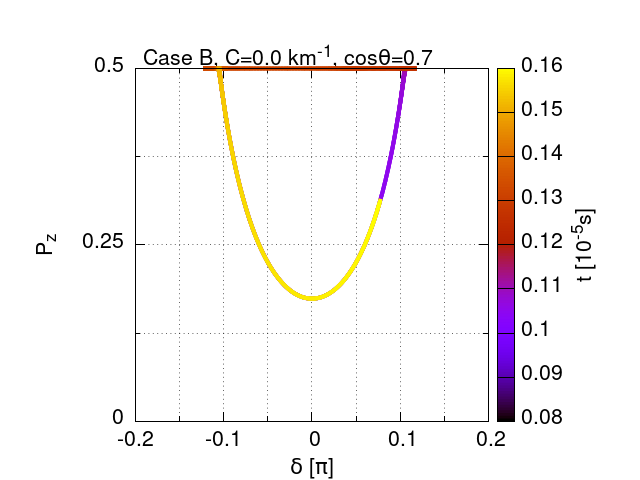}\\
\includegraphics[width=0.95\linewidth]{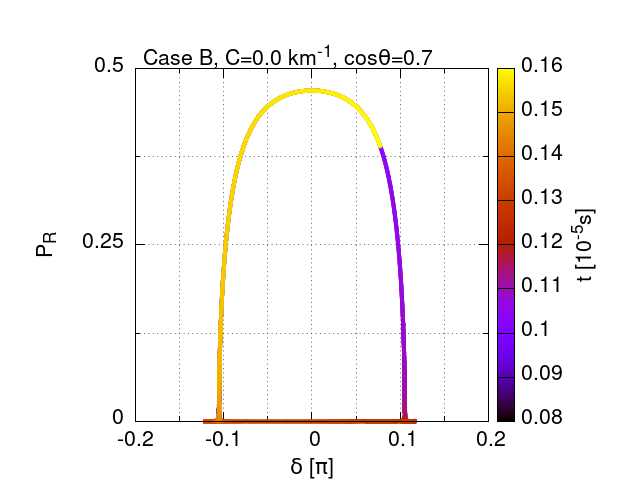}
\caption{%%
Top: The time evolution of neutrino polarization vector ($\cos\theta=0.7$) on the $\delta$-$P_{z}$ plane from $0.08\times10^{-5}$\,s to $0.16\times10^{-5}$\,s. Here, we set $C=0\,\mathrm{km}^{-1}$. At $t=0.08\times10^{-5}$\,s, $\delta$ is close to $-0.1\pi$. The polarization vector moves counterclockwise along the track.
Bottom: The time evolution of polarization vector on the $\delta$-$P_{R}$ plane. The polarization vector moves clockwise along the track.
}
\label{fig:d-z01_C0_caseb}
\end{figure}

\begin{figure}[t]
\includegraphics[width=0.95\linewidth]{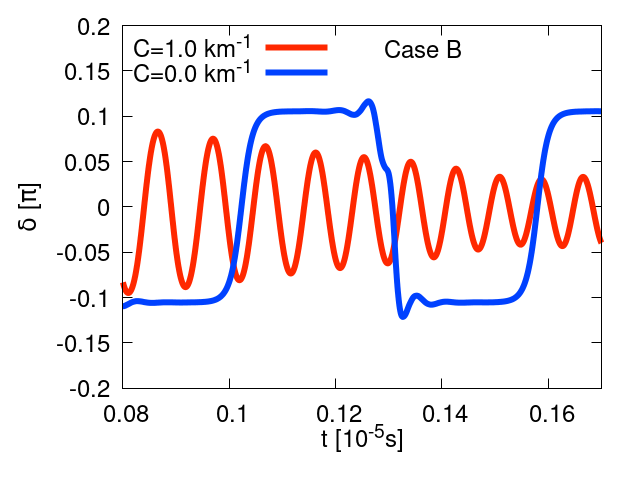}\\
\includegraphics[width=0.95\linewidth]{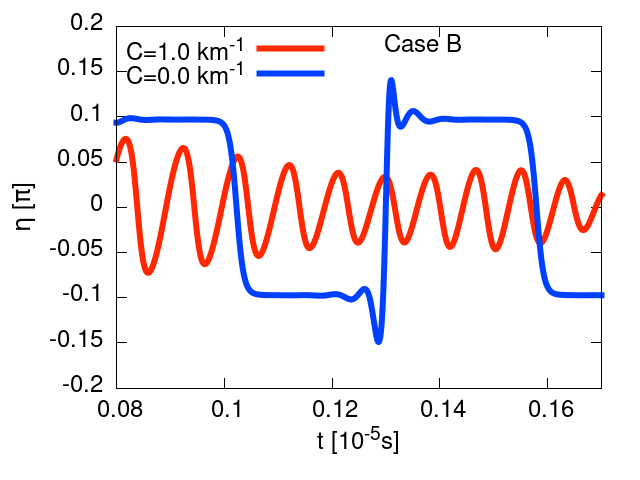}
\caption{%%
Top: The time evolution of $\delta$ at $\cos\theta=0.7$ in the limit cycle phase. 
Bottom: The time evolution of $\eta=P_\phi -\langle P \rangle_{\phi}$ at $\cos\theta=0.7$ in the limit cycle phase. 
The blue (red) line shows the case of $C=0\,\mathrm{km}^{-1}$ ($C=1\,\mathrm{km}^{-1}$).
}
\label{fig:t-delta_zoom_caseb}
\end{figure}

\begin{figure}[t]
\includegraphics[width=0.95\linewidth]{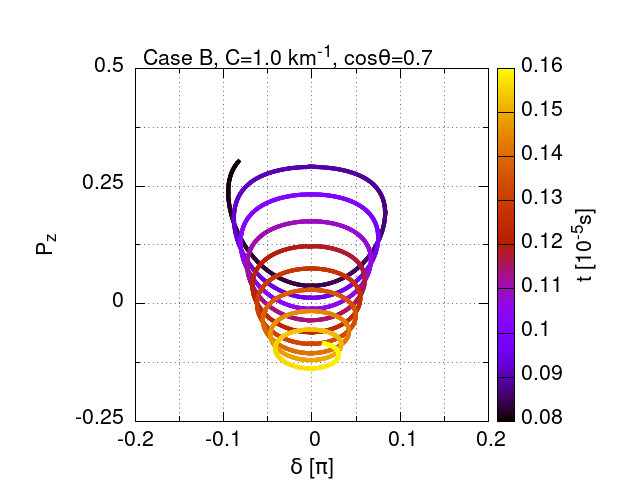}\\
\includegraphics[width=0.95\linewidth]{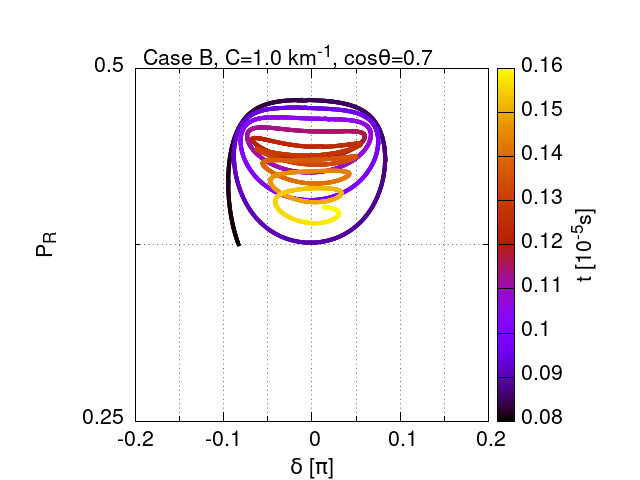}
\caption{%%
Same to Fig.~\ref{fig:d-z01_C0_caseb} but here, we consider the collision term ($C=1\,\mathrm{km}^{-1}$).
}
\label{fig:d-z01_C1_caseb}
\end{figure}

%\begin{figure}[t]
%\includegraphics[width=0.95\linewidth]{spectrogram_caseb.png}\\
%\caption{%%
%The time evolution of the spectrogram of $\delta$ at $\cos\theta=0.7$. The unit of the color bar is dB.
%}
%\label{fig:spectrogram_caseb}
%\end{figure}

\begin{figure}[t]
\includegraphics[width=0.95\linewidth]{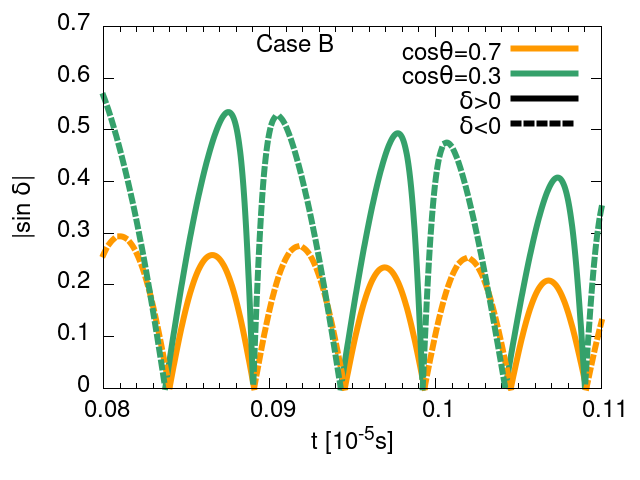}
\caption{
The time evolution of $|\sin\delta|$ at $\cos\theta=0.7$ (orange) and $\cos\theta=0.3$ (green).
For positive and negative $\delta$, solid and dashed curve are used, respectively.
}
\label{fig:t-abs_sin_delta_caseb}
\end{figure}

Here, we define the limit cycle phase as the period from the linear evolution phase until $\average{P_{ex}}$ settles down to its equilibrium value. Fig.~\ref{fig:t-Pex_zoom_casseb} shows an enlarge view of Fig.~\ref{fig:caseB-t-Pex} focusing on the flavor conversions in the limit cycle phase. In the case without collision terms (blue curve), flavor conversions become periodic. On the other hand, in the case with collision term, the amplitude of the flavor conversion becomes gradually smaller and the transition probability is enhanced by collision effect (red curve). In the limit cycle phase, the collision effect is no longer negligible in flavor conversions. We study the collision effect on the motion of neutrino polarization vectors in the limit cycle phase. The equation of motion of $\mathbf{P}$ in Eq.~(\ref{eq:EOM neutrino vector}) is decomposed by
\begin{align}
\label{eq:time evolution Pz in the limit cycle phase}
    \frac{\mathrm{d}}{\mathrm{d}t}P_{z}&=\;\; H_{R}P_{R}\sin\delta-C\left(
    P_{z}-\average{P_{z}}
    \right),\\
\label{eq:time evolution Pr in the limit cycle phase}
    \frac{\mathrm{d}}{\mathrm{d}t}P_{R}&=-H_{R}P_{z}\sin\delta-C\left(
    P_{R}-\average{P}_{R}\cos\eta
    \right),\\
P_R \frac{{\rm d} P_\phi}{{\rm d} t} &=
-H_R P_z \cos\delta 
-C\langle P\rangle_R\sin\eta+P_R H_z
\label{eq:time evolution Pp in the limit cycle phase}
\end{align}
where $H_{R}$ and $\delta$ are given in Eq.~(\ref{eq:phase difference}), and $\eta=P_\phi -\langle P \rangle_\phi$.
Note that we first calculate average polarization $\langle P_x \rangle$ and $\langle P_y \rangle$, and then obtain 
$\average{P}_R=\sqrt{\average{P_x}^2+\average{P_y}^2}$ and 
$\langle P \rangle_\phi = \tan^{-1}\left(\average{P_y} / \average{P_x}\right)$. In general, 
$\average{P}_R \neq \average{P_R}$
and 
$\average{P}_\phi \neq \average{P_\phi}$, where the right hand sides are the average of $P_R$ and $P_\phi$, respectively.

Fig.~\ref{fig:d-z01_C0_caseb} shows the time evolution of $P_{z}$ and $P_{R}$ at $\cos\theta=0.7$ without the collision ($C=0\,\mathrm{km}^{-1}$) during $t$ in $0.08\text{--}0.16$ in the unit of $10^{-5}\,\mathrm{s}$, respectively. Without collision terms, the evolution tracks of polarization vectors in Fig.~\ref{fig:d-z01_C0_caseb} are closed. The blue curve in the top panel of Fig.~\ref{fig:t-delta_zoom_caseb} shows the evolution of $\delta$ without collision effect. At first, the value of $\delta$ is approximately $-0.1\pi$ and almost constant. The value of $\delta$ increases dramatically around $t=0.1\times10^{-5}$\,s. According to the first terms on the right hand sides of Eqs.~(\ref{eq:time evolution Pz in the limit cycle phase}) and (\ref{eq:time evolution Pr in the limit cycle phase}), the value of $P_{z}$ ($P_{R}$) decreases (increases) when the $\delta$ is negative. In the case of positive $\delta$, the value of $P_{z}$ ($P_{R}$) increases (decreases). The perpendicular component is negligible ($P_{R}\sim0$) and the $z$-component is constant ($P_{z}\sim 0.5$) at $t=(0.12\text{--}0.14)\times10^{-5}$\,s irrespective of decreasing $\delta$. In such evolution phase, the small $H_{R}$ suppresses the time evolution of $P_{z}$ and $P_{R}$. Therefore, the polarization vector moves counterclockwise (clockwise) in 
Top panel (Bottom panel) of Fig.~\ref{fig:d-z01_C0_caseb}.

In the case with neutrino scatterings ($C=1\,\mathrm{km}^{-1}$), the evolution track of neutrino polarization vector is no longer a closed orbit but a ``limit cycle" on the phase space because of the collision terms in Eqs.(\ref{eq:time evolution Pz in the limit cycle phase}) and (\ref{eq:time evolution Pr in the limit cycle phase}). The spiral motion of $P_{z}$ and $P_R$ at $\cos\theta=0.7$ is shown in the top and bottom panels of Fig.~\ref{fig:d-z01_C1_caseb}, respectively.
 The value of $\delta$ is negative at $t=0.08\times10^{-5}$\,s (see the red curve in the top panel of  Fig.~\ref{fig:t-delta_zoom_caseb}). By the first term on the right hand side of Eq.~(\ref{eq:time evolution Pz in the limit cycle phase}) (Eq.~\eqref{eq:time evolution Pr in the limit cycle phase}), the value of $P_{z}$ ($P_{R}$)  decreases (increases) at first.
On the other hand, in more later phase, $\delta$ becomes positive and $P_{z}$ ($P_{R}$)  increases (decreases).

The ranges of $P_{z}$ and  $P_{R}$ changing in one cycle gradually decrease with each cycle.
As shown in the red curve of the top panel of Fig.~\ref{fig:t-delta_zoom_caseb}, the oscillation amplitude of $\delta$ is decreasing and converging to zero.
In the bottom panel, the evolution of $\eta$ is also shown.
Comparing the top and bottom panels, we found $\delta\sim - \eta$.
Since the collision tries to align the polarization and decreases $|\eta|=|P_\phi-\langle P\rangle_\phi |$, $|\delta|\sim|\eta|$ is also expected to become smaller as time passes.
%Such feature is related to the increasing characteristic frequency, $f$, in the spectrogram of $\delta$ in Fig.~\ref{fig:spectrogram_caseb} where the spetctrogram is short-time Fourier transformation in each time.
%The increasing $f$ means that the period of each cycle becomes smaller.
After such synchronization between $\mathbf{P_{R}}$ and $\mathbf{H_{R}}$, the first terms on the right hand side of Eqs.~\eqref{eq:time evolution Pz in the limit cycle phase} and \eqref{eq:time evolution Pr in the limit cycle phase} are negligible and the transition probability $\average{P_{ex}}$ no longer changes, which results in the end of the limit cycle phase.
 
 One of the most interesting features of this model is that the mean value of $P_z$ becomes smaller as time passes (e.g., top of Fig.~\ref{fig:d-z01_C1_caseb}).
 Fig.~\ref{fig:t-abs_sin_delta_caseb} shows $|\sin \delta|$. It is similar to Fig.~\ref{fig:t-delta_zoom_caseb}, but we can compare positive $\delta$ and negative $\delta$ easily. In the case of $\cos\theta=0.7$, the averaged $|\sin \delta|$ for negative $\delta$ is larger than that of positive $\delta$ in one cycle. In addition, the period with negative $\delta$ is slightly longer than the positive part.
 From Eq.~\eqref{eq:time evolution Pz in the limit cycle phase},
 this imbalance of positive and negative $\delta$ leads to the gradual decrease of $P_z$.
 Note that this does not happen for all $\theta$.
In the case of $\cos\theta=0.3$, on the other hand, the positive part is slightly larger than the positive part.
This excess slightly increases the mean $P_z$ (see the bottom of Fig.~\ref{fig:perp relaxation_caseb}).

This imbalance comes from the highly non-linear dynamics of the partially synchronized oscillators,
and it is difficult to identify the mechanism to produce that. Here we show a hypothesis of the possible origin, focusing on 
the impact of the collision term on $\delta=P_\phi-H_\phi$. Note that other terms may also cause the imbalance.
First we extract the impact of the collision term on the phase of the polarization vector:
$\left.\frac{{\rm d}P_\phi}{{\rm d}t}\right|_{\rm coll}= \frac{{\rm d}P_\phi}{{\rm d}t} - \left.\frac{{\rm d}P_\phi}{{\rm d}t}\right|_{C=0}$, where $\left.\frac{{\rm d}P_\phi}{{\rm d}t}\right|_{C=0}$ is obtained by substituting $C=0$ in Eq.~\eqref{eq:time evolution Pp in the limit cycle phase}. Similarly, we consider the collisional part of $\delta$: $\left.\frac{{\rm d}\delta}{{\rm d}t}\right|_{\rm coll}= \frac{{\rm d}\delta}{{\rm d}t} - \left.\frac{{\rm d}\delta}{{\rm d}t}\right|_{C=0}$ .
%though we cannot explicitly write down the equation.
%The term 
This term could be well approximated as 
\begin{align}
\label{eq:Time evolution of delta}
\left.\frac{{\rm d}\delta}{{\rm d}t}\right|_{\rm coll} 
&=
 \left.\frac{{\rm d}P_\phi}{{\rm d}t}\right|_{\rm coll}
-\left.\frac{{\rm d}H_\phi}{{\rm d}t}\right|_{\rm coll}
&\sim-C\left(\frac{\average{P}_R}{P_R}\right)\sin\eta,
\end{align}
where we ignore the effect of the collision term in $\frac{{\rm d} H_\phi}{{\rm d} t}$ following Appendix~\ref{sec:derivation of time evolution of delta},
and the collision term in Eq.~\eqref{eq:time evolution Pp in the limit cycle phase} is extracted.

\begin{figure}[htbp]
\includegraphics[width=0.8\linewidth]{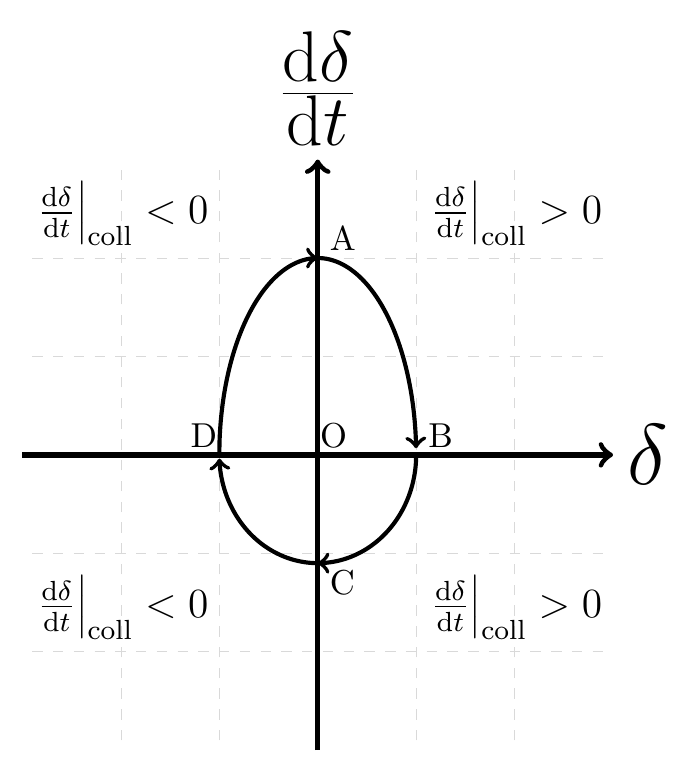}
\caption{%%
Schematic diagram of the evolution of $\delta$ and $\frac{{\rm d}\delta}{{\rm d} t}$ .
}
\label{fig:schematic}
\end{figure}

We demonstrate how the collision term violates the time symmetry using the schematic diagram of Fig.~\ref{fig:schematic}.
Near $t=0.0945\times 10^{-5}\,{\rm s}$ in Fig.~\ref{fig:t-abs_sin_delta_caseb},
$\delta$ of $\cos\theta=0.7$ changes its sign from negative to positive, i.e., $\frac{{\rm d}\delta}{{\rm d}t}>0$ (point A in the diagram).
The collision term would decelerate $\delta$ when $t < 0.0945\times 10^{-5}\,{\rm s}$ and $\delta <0$ since
the collision term, $\left.\frac{{\rm d}\delta}{{\rm d}t}\right|_{\rm coll}$, is proportional to $-C \sin\eta \sim C\sin\delta < 0$ ( here $\eta\sim -\delta$. see Fig.~\ref{fig:t-delta_zoom_caseb}).
Now $\frac{{\rm d}\delta}{{\rm d}t}$ is positive, and $\left.\frac{{\rm d}\delta}{{\rm d}t}\right|_{\rm coll}$ is negative, then the evolution of $\delta$ would be decelerated by the collision (curve of D to A in the diagram).
On the other hand, when $t > 0.0945\times 10^{-5}\,{\rm s}$ and  $\delta >0$,
the collision term accelerate the evolution of $\delta$ since $\frac{{\rm d}\delta}{{\rm d}t}>0$ and 
$\left.\frac{{\rm d}\delta}{{\rm d}t}\right|_{\rm coll}$ are both positive (Curve A-B in the diagram).
These two effects make the duration in negative $\delta$ longer and positive $\delta$ shorter.
Caveat that at $t=0.099\times 10^{-5}\,{\rm s}$, the collision term, vise versa, makes 
the duration in negative $\delta$ shorter and the duration in positive $\delta$ longer.
Here $\frac{{\rm d}\delta}{{\rm d}t}<0$, and $\left.\frac{{\rm d}\delta}{{\rm d}t}\right|_{\rm coll}>0$ ($<0$) for positive (negative) $\delta$ in Curve B-C (C-D).
We need more careful analysis for quantitative argument, which we keep in a future study as well as the possibility of other origin.

\subsection{Relaxation phase}
\label{sec:Relaxation phase}

\begin{figure}[t]
\includegraphics[width=0.95\linewidth]{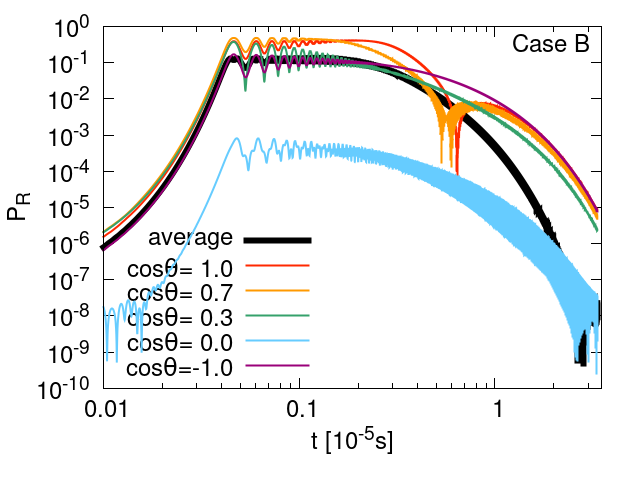}\\
\includegraphics[width=0.95\linewidth]{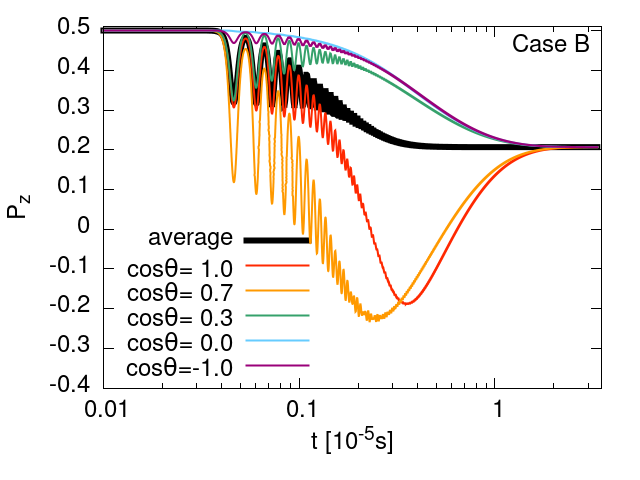}
\caption{%%
Top: The time evolution of perpendicular component of neutrino polarization vectors at different angles (color curves), $P_R$, and that of angle averaged one (black), $\langle P\rangle_{R}$.
Bottom: The time evolution of $P_{z}$ at different angles and that of the angle averaged $z$-component $\average{P_{z}}$.
The red, orange, green, cyan, and purple curves show the case of $\cos\theta=1,0.7,0.3,0$ and $-1$, respectively.
}
\label{fig:perp relaxation_caseb}
\end{figure}

As implied in Eq.~(\ref{eq:mean square polarization vectors}), the distributions of neutrinos affected by fast flavor conversions become isotropic in the relaxation phase ($t>C^{-1}=0.33\times 10^{-5}$\,s). Up to the limit cycle phase, the fast flavor conversions are enhanced by collision effects and the transition probability $\average{P_{ex}}$ does not change anymore as shown in Fig.~\ref{fig:caseB-t-Pex}. In the relaxation phase, the collision terms force all of polarization vectors to face $z$-axis keeping the value of $\average{P_{ex}}$.

\begin{figure}[htbp]
\includegraphics[width=0.9\linewidth]{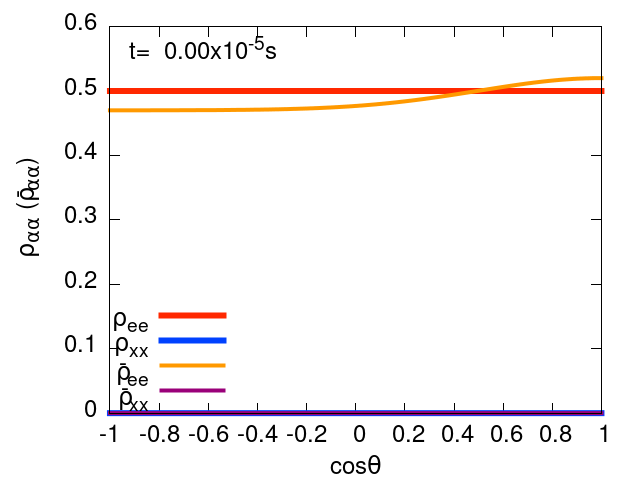}\\
\includegraphics[width=0.9\linewidth]{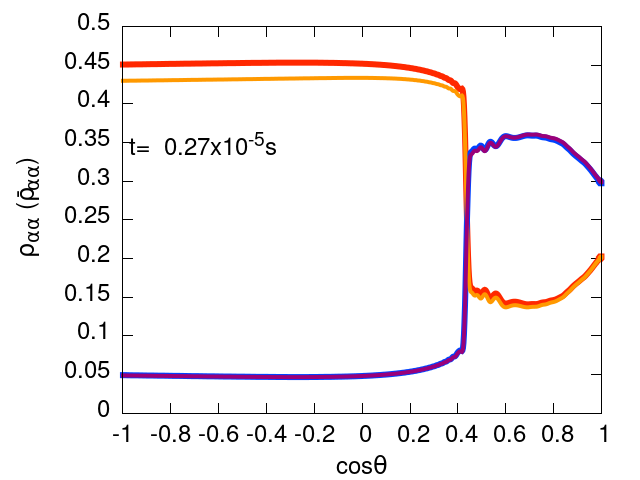}\\
\includegraphics[width=0.9\linewidth]{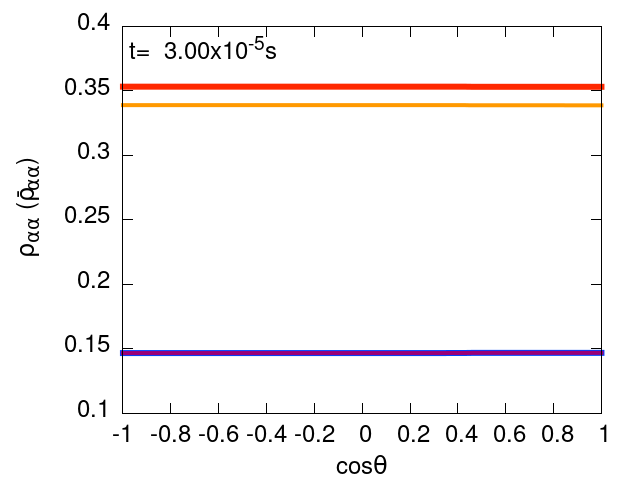}
\caption{%%
Time snapshots of neutrino spectra.
The initial ($t=0$\,s), after limit cycle ($t=0.27\times 10^{-5}$\,s), after the relaxation ($t=3.00\times 10^{-5}$\,s) snapshots are taken, respectively.
The distribution of $\rho_{ee}$, $\rho_{xx}$, $\bar{\rho}_{ee}$, and $\bar{\rho}_{xx}$  
correspond to red, blue, orange, and purple curves, respectively.
}
\label{fig:spec relaxation_caseb}
\end{figure}

The top panel of Fig.~\ref{fig:perp relaxation_caseb} shows the time evolution of $P_{R}$ in different angles and that of $\average{P}_{R}$. Owing to the coupling of neutrino-neutrino interactions with neutrino scatterings, the value of $P_{R}$ increases and saturates around $t\sim 0.1\times 10^{-5}$\,s.
Note that the time of the first peak would depend on the initial perturbation. We impose $P_{R}\sim 10^{-8}$, initially.
After the saturation, all of polarization vector starts to be parallel to the %
$z$-axis and the values of $P_{R}$ are reduced to zero. The perpendicular component of $\cos\theta=0$ (cyan curve) is small and the polarization vector always points near the $z$-axis. From Eqs.(\ref{eq:EOM neutrino vector}) and (\ref{eq:vectors}), equation of motions of $\mathbf{D_{0}}$ is described by 
\begin{equation}
\label{eq:EOM of D0}
    \frac{\mathrm{d}}{\mathrm{d}t}\mathbf{D_{0}}=\omega\left(
    \average{\mathbf{P}}+\average{\mathbf{\bar{P}}}
    \right)\times\mathbf{B}.
\end{equation}
%The above equation of motion indicates that, in the case of a small mixing angle ($\theta_{v} \ll 1$) and a small vacuum frequency ($\omega t \ll 1$), both $\mathbf{B}$ and $\mathbf{D_{0}}$ become parallel to the $z$-axis during the flavor conversions. Then, $\omega\mathbf{B}+\mu^{\prime}\mathbf{D_{0}}$ in Eq.~(\ref{eq:EOM neutrino vector}) is also parallel to the $z$-axis, so that the $P_{R}$ is almost negligible at $\cos\theta=0$.
The above equation of motion indicates that, in the case of a small vacuum frequency ($\omega t \ll 1$), $\mathbf{D_{0}}$ is almost conserved during the flavor conversions and parallel to the $z$-axis. In addition, $\mathbf{B}$ is also directed to the $z$-axis in the case of a small mixing angle ($\theta_{\mathrm{v}} \ll 1$). Then, $\omega\mathbf{B}+\mu^{\prime}\mathbf{D_{0}}$ in Eq.~(\ref{eq:EOM neutrino vector}) is also parallel to the $z$-axis, so that the $P_{R}$ is almost negligible at $\cos\theta=0$. The evolution of $P_{z}$ in different angles and that of $\average{P_{z}}$ are shown in the bottom panel of Fig.~\ref{fig:perp relaxation_caseb}. The fast flavor conversions increase the transition probability and induce different values of $P_{z}$ depending on the scattering angle $\theta$ until the end of the limit cycle $t\sim C^{-1}=0.33\times 10^{-5}$\,s. The flavor conversions in the limit cycle phase are especially enhanced at $\cos\theta=0.7$ (orange curve) and $\cos\theta=1$ (red curve) in the bottom panel of Fig.~\ref{fig:perp relaxation_caseb}. Such enhancement of flavor conversions in forward scattered neutrinos is also confirmed in the middle panel of Fig.\ref{fig:spec relaxation_caseb}. However, after the limit cycle, the values of $P_{z}$ in different angles (color curves) converge on the value of $\average{P_{z}}$ (black curve). Then, the fast flavor conversions of neutrinos have finished when $P_{z}=\average{P_{z}}$ is satisfied in all of the scattering angles. Such behaviors of $P_{z}$ can be understood by the equation of motion of $P_{z}$ in the relaxation phase. Eq.(\ref{eq:transition probability}) shows that the $\average{P_{z}}$ is related to the transition probability as  $\average{P_{ex}}=0.5-\average{P_{z}}$ in our numerical setup. As discussed in Sec. \ref{sec:Limit cycle phase}, the time evolution of $\average{P_{ex}}$ reaches an equilibrium in the limit cycle phase owing to the synchronization between polarization vectors of neutrino density matrix and that of Hamiltonian ($\mathbf{P_{R}}\parallel\mathbf{H_{R}}$). Therefore, the value of $\delta$ is negligible and the value of $\average{P_{z}}$ becomes constant before the relaxation phase. Then, the equation of motion of $P_{z}$ in the relaxation phase is derived from Eq.~(\ref{eq:time evolution Pz in the limit cycle phase}),
\begin{equation}
    \frac{\mathrm{d}}{\mathrm{d}t}P_{z}\sim-C\left( P_{z}-\average{P_{z}}
    \right),
\end{equation}
which explains the relaxation of $P_{z}$ to $\average{P_{z}}$ in all scattering angles at $t> C^{-1}=0.33\times 10^{-5}$\,s as shown in the bottom panel of Fig.\ref{fig:perp relaxation_caseb}. Here, we discuss the relaxation of polarization vectors in neutrino sector, but the case of antineutrinos are similar. In the case of antineutrinos, the polarization vector $\mathbf{\bar{P}}$ is relaxed to $(0,0,\average{\bar{P}})$ irrespective of the scattering angle $\theta$.

The angular distributions of neutrinos after fast flavor conversions are shown in Fig.~\ref{fig:spec relaxation_caseb} bottom panel. As implied in Eq.~(\ref{eq:mean square polarization vectors}), the angular dependence disappears in the distribution of neutrinos. The density matrices of neutrinos (antineutrinos) become diagonal because of the negligible $P_{R}$($\bar{P}_{R}$) and the finite $P_{z}$($\bar{P}_{z}$). 
The middle panel is taken at $t=0.27\times 10^{-5}$\,s.
Up to the limit cycle, the values of $\rho_{ee}$ and $\bar{\rho}_{ee}$ depend on the scattering angle \cite{Shalgar:2020wcx}.
It looks similar to the stationary solution without collision \cite{Xiong:2021dex}.
However, such angular dependence disappear during the relaxation phase and the neutrino distribution becomes isotropic because of the collision effect. In Fig.~\ref{fig:spec relaxation_caseb} bottom panel, the value of $\rho_{xx}$ (blue curve) is equal to that of $\bar{\rho}_{xx}$ (purple curve). Such correspondence can be explained by a conservation law below
\begin{equation}
\label{eq:conservation law D0}
    %\mathbf{B}\cdot\frac{\mathrm{d}}{\mathrm{d}t}\mathbf{D_{0}}=0,
    \mathbf{B}\cdot\mathbf{D_{0}}=\mathrm{const.}\,.
\end{equation}
The above equation is derived from Eq.~(\ref{eq:EOM of D0}). In the case of small vacuum mixing angle, $\average{P_{z}}-\average{\bar{P}_{z}}$ is almost conserved according to Eq.~(\ref{eq:conservation law D0}). Furthermore the $\average{\mathrm{Tr}\rho}$ and $\average{\mathrm{Tr}\bar{\rho}}$ are time-independent quantities. Therefore, we can show that $\average{\rho_{xx}}$ is almost equal to $\average{\bar{\rho}_{xx}}$ at any time:
\begin{equation}
\begin{split}
    \average{\rho_{xx}}-\average{\bar{\rho}_{xx}}&=\frac{\average{\mathrm{Tr}\rho}}{2}-\frac{\average{\mathrm{Tr}\bar{\rho}}}{2}-\frac{\average{P_{z}}}{2}+\frac{\average{\bar{P}_{z}}}{2}\\
    &\sim\left(
    \average{\rho_{xx}}-\average{\bar{\rho}_{xx}}
    \right)(t=0)\\
    &=0.
    \end{split}
\end{equation}
After the relaxation phase, neutrino distribution becomes isotropic, so that $\rho_{xx}\sim\bar{\rho}_{xx}$ is satisfied in the end of the calculation as shown in Fig.~\ref{fig:spec relaxation_caseb}.

Note that the appearance of the relaxation phase depends on the situation.
The ELN crossing, considered here, typically appears above the neutrino sphere \citep{Nagakura2021a}, and it means the opacity is less than one, i.e., $t < C^{-1}$. 
On the other hand, the ELN crossing in proto-neutron stars,
opacity is larger than one, and  the relaxation phase should be considered.
In any case, the discussion here would be helpful to understand the role of the collision term.

\subsection{Dependence of neutrino scattering terms}
%\label{sec:}

\begin{figure}[htbp]
\includegraphics[width=0.9\linewidth]{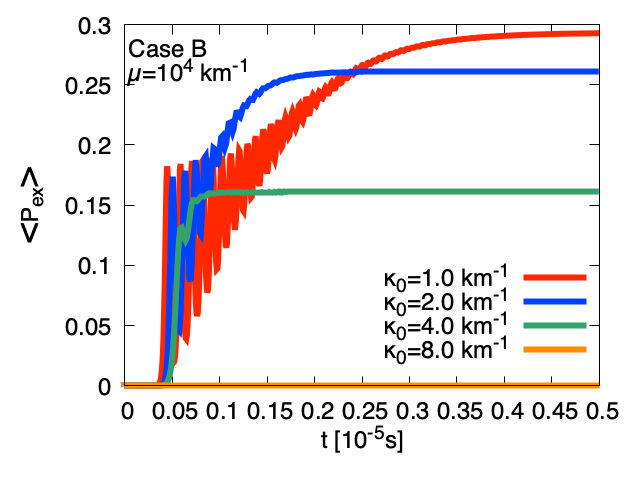}\\
\includegraphics[width=0.9\linewidth]{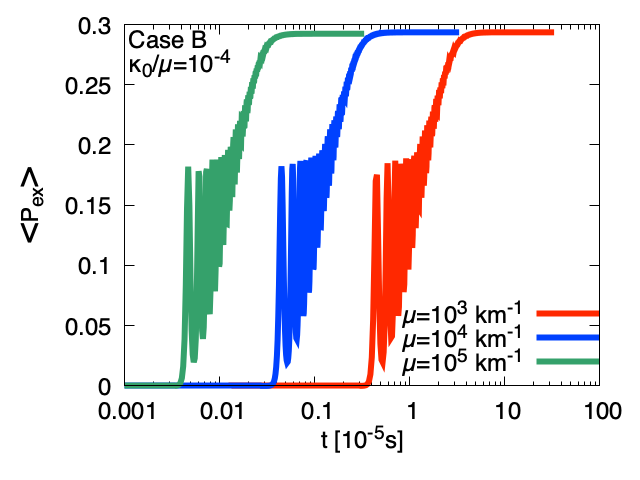}
\caption{%%
Top: The time evolution of Eq.(\ref{eq:transition probability general}) with NC collision terms in Eqs.~(\ref{eq:collision rho martin}) and (\ref{eq:collision rhob martin}) with different values of $\kappa_{0}$. Bottom: The results with different values of $\mu$ with $\kappa_{0}/\mu=10^{-4}$. Here, we finished the calculations of $\mu=10^{n}\,\mathrm{km}^{-1}$ at $t=3.3\times10^{-1-n}$ s ($n=3,4,5$).
}
\label{fig:collision martin2021}
\end{figure}

\begin{figure}[htbp]
\includegraphics[width=0.9\linewidth]{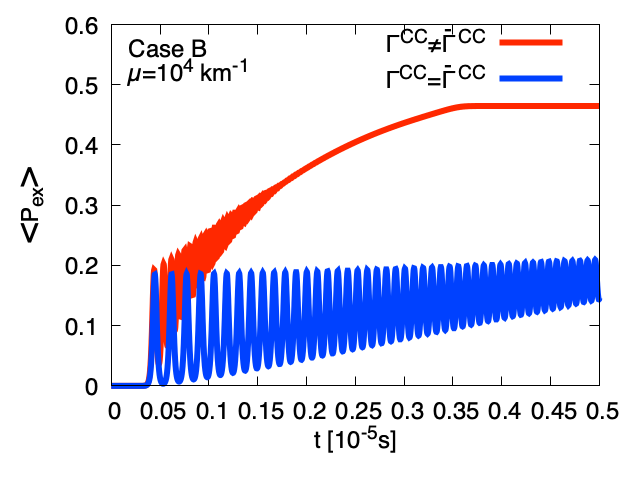}\\
\includegraphics[width=0.9\linewidth]{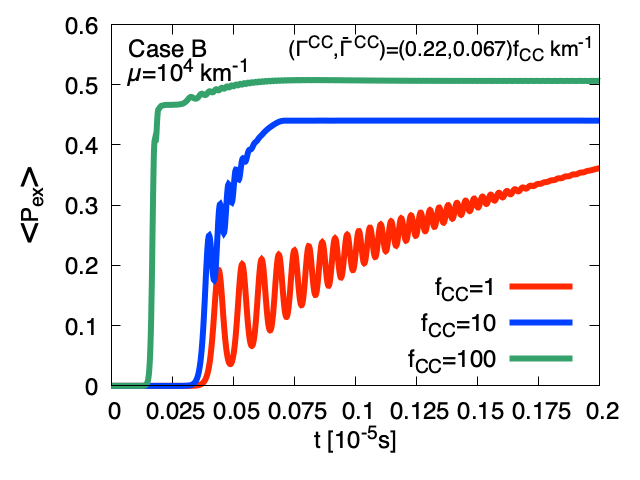}
\caption{%%
Top: The time evolution of Eq.(\ref{eq:transition probability general}) with CC collision terms in Eqs.~(\ref{eq:collision rho johns}) and (\ref{eq:collision rhob johns}). The red curve shows the result of asymmetric case, i.e., $\Gamma^{\rm CC}=0.22\, \mathrm{km}^{-1}$ and  $\bar{\Gamma}^{\rm CC}=0.067\,\mathrm{km}^{-1}$.
The blue curve shows the result of the symmetric case, $\Gamma^{\rm CC}=\bar{\Gamma}^{\rm CC}=0.22\,\mathrm{km}^{-1}$. Bottom: The results with different strengths of the CC collision terms, $\Gamma^{\rm CC}=0.22f_{\rm CC}\, \mathrm{km}^{-1}$ and $\bar{\Gamma}^{\rm CC}=0.067f_{\rm CC}\, \mathrm{km}^{-1}$, where $f_{\rm CC}=1, 10,$ and $100$.
}
\label{fig:collision johns2021}
\end{figure}

In the previous section, we confirm that the collision terms of Eqs.~(\ref{eq:collision rho}) and (\ref{eq:collision rhob}) enhance fast neutrino flavor conversions and make neutrino angular distributions isotropic. In this section, we calculate angle averaged transition probabilities by using collision terms of neutrino scatterings in Refs.~\cite{Martin:2021xyl,Johns:2021arXiv210411369J}. We use the same numerical setup as that of the previous section except for the collision term.

First, we consider the collision terms of neutrino scatterings in neutral-current (NC) reactions.
For the updated NC reaction, we employ the elastic neutrino-nucleon collision in Ref.~\cite{Martin:2021xyl},
\begin{align}
    C^{\mathrm{NC}}[\rho]=&-\kappa_{0}\rho(\cos\theta)\nonumber\\
    &+\frac{1}{2}\int^{1}_{-1}\mathrm{d}\cos\theta^{\prime}\left(\kappa_{0}-\frac{\kappa_{1}}{3}\cos\theta\cos\theta^{\prime}\right)\rho(\cos\theta^{\prime}), \label{eq:collision rho martin}\\
    \bar{C}^{\mathrm{NC}}[\bar{\rho}]=&-\kappa_{0}\bar{\rho}(\cos\theta)\nonumber\\
    &+\frac{1}{2}\int^{1}_{-1}\mathrm{d}\cos\theta^{\prime}\left(\kappa_{0}-\frac{\kappa_{1}}{3}\cos\theta\cos\theta^{\prime}\right)\bar{\rho}(\cos\theta^{\prime}),\label{eq:collision rhob martin}
\end{align}
where we fix $\kappa_{1}/\kappa_{0}=0.5$ as in Ref.\cite{Martin:2021xyl}. The gain terms in the above NC collisions depend on the neutrino scattering angle $\theta$. The collision terms in Eqs.~(\ref{eq:collision rho}) and (\ref{eq:collision rhob}) are reproduced, when $\kappa_{1}=0\,\mathrm{km}^{-1}$ and $\kappa_{0}=C$.

The top panel of Fig.~\ref{fig:collision martin2021} shows the evolution of the transition probability $\average{P_{ex}}$ in Eq.~(\ref{eq:transition probability general}) with different values of $\kappa_{0}$. In the case of $\kappa_{0}=1\,\mathrm{km}^{-1}$ (red curve), the flavor conversion is well enhanced by the collision effect and the result is almost identical to that of $C=1\,\mathrm{km}^{-1}$ in Fig.~\ref{fig:caseB-t-Pex}. This correspondence suggests the small contribution of $\kappa_{1}$ to the NC scattering. As shown in the blue, green, and orange curves in Fig.~\ref{fig:collision martin2021}, the flavor conversions are more suppressed in the large value of $\kappa_{0}$. Any flavor conversion does not appear in $\kappa_{0}\geq 8\,\mathrm{km}^{-1}$, where all of the polarization vectors are directed to the $z$-axis by strong collision terms so that the instability for fast flavor conversions does not grow up sufficiently. Nothing is happening in the neutrino sector, but the distribution of electron antineutrino becomes isotropic following the classical Boltzmann equation without neutrino oscillations. Such damping effect of a large $\kappa_{0}$ is consistent with the result of Ref.~\cite{Martin:2021xyl}.

The bottom panel of Fig.~\ref{fig:collision martin2021} shows the results with different values of $\mu$ while maintaining the ratio, $\kappa_{0}/\mu=10^{-4}$. The transition probabilities scale to $\mu^{-1}$ and the flavor conversion is raised even with the large collision parameter, $\kappa_{0}=10\,\mathrm{km}^{-1}$ (green curve). Therefore, the collision effect can be characterized by the ratio, $\kappa_{0}/\mu$. It seems that a small collision term, $\kappa_{0}\leq \tau^{-1}_{\rm{osc}}\propto\mu$ is required to enhance the flavor conversions, where $\tau_{\rm{osc}}$ is a characteristic oscillation time proportional to $\mu^{-1}$.

%This oscillation time might correspond to $\tau_{\rm bipolar}$  \cite{Hannestad:2006nj} in the framework of fast flavor conversions.

Next, we consider charged-current (CC) reactions.
For the CC scatterings, we study the effect of neutrino electron scatterings following the collision term in Ref.~\cite{Johns:2021arXiv210411369J},
\begin{align}
    %C^{\mathrm{CC}}[\rho]=&\Gamma_{CC}\left(
    %\begin{array}{cc}
    % \average{\rho_{ee}}-\rho_{ee}    & -\frac{\rho_{ex}}{2} \\
    %  -\frac{\rho_{xe}}{2}   & 0
    %\end{array}
    %\right),
    C^{\mathrm{\rm CC}}[\rho]=&-\Gamma^{\rm CC}\left(
    \begin{array}{cc}
     \rho_{ee}    & \frac{\rho_{ex}}{2} \\
      \frac{\rho_{xe}}{2}   & 0
    \end{array}
    \right)+\Gamma^{\rm CC}\left(
    \begin{array}{cc}
     \average{\rho_{ee}}    & 0\\
      0  & 0
    \end{array}
    \right),
    \label{eq:collision rho johns}\\
    \bar{C}^{\mathrm{\rm CC}}[\bar{\rho}]=&-\bar{\Gamma}^{\rm CC}\left(
    \begin{array}{cc}
   \bar{\rho}_{ee}    & \frac{\bar{\rho}_{ex}}{2} \\
    \frac{\bar{\rho}_{xe}}{2}   & 0
    \end{array}
    \right)+\bar{\Gamma}^{\rm CC}\left(
    \begin{array}{cc}
     \average{\bar{\rho}_{ee}}  & 0\\
      0   & 0
    \end{array}
    \right),
    \label{eq:collision rhob johns}
\end{align}
where $\Gamma^{\rm CC}=1/\lambda_{\nu_{e}e}$ and $\bar{\Gamma}^{\rm CC}=1/\lambda_{\bar{\nu}_{e}e}$ are calculated from the rates of the electron scatterings (see Eq.~($11$) in Ref.~\cite{Johns:2021arXiv210411369J}).
%We mention that Eq.~(\ref{eq:transition probability}) is not satisfied with the numerical setup with the above CC collision. The transition probability $\average{P_{ex}}$ is calculated from the definition in Ref.~\cite{Shalgar:2020wcx}.}
The red curve in the top panel of Fig.~\ref{fig:collision johns2021} shows the transition probability $\average{P_{ex}}$ in Eq.~(\ref{eq:transition probability general}) for the typical values at a post-bounce near the proto-neutron star, $\Gamma^{\rm CC}=0.22\,\mathrm{km}^{-1}$ and $\bar{\Gamma}^{\rm CC}=0.067\,\mathrm{km}^{-1}$ for $E=50$\,MeV neutrinos. The transition probability increases due to the collision effect as in Fig.~\ref{fig:caseB-t-Pex}. The equilibrium value of transition probability is larger than that of the NC scattering because of the large asymmetry, $\Gamma^{\rm CC}\neq\bar{\Gamma}^{\rm CC}$. On the other hand, the flavor conversion does not equilibrate immediately for the symmetric collision parameter, $\Gamma^{\rm CC}=\bar{\Gamma}^{\rm CC}=0.22\,\mathrm{km}^{-1}$, as shown in the blue curve in the top panel of Fig.~\ref{fig:collision johns2021}. Such property between $\Gamma^{\rm CC}$ and $\bar{\Gamma}^{\rm CC}$ is also reported in Ref.~\cite{Johns:2021arXiv210411369J}. Our results clarify the role of the asymmetry on the CC scatterings.

In the symmetric case, when $\Gamma^{\rm CC}=\bar{\Gamma}^{\rm CC}\geq 8\, \mathrm{km}^{-1}$, the flavor conversions are strongly suppressed, as in the NC scattering. On the other hand, as shown in the bottom panel of Fig.~\ref{fig:collision johns2021}, the flavor conversions in the asymmetric case are not suppressed even in the large collision terms ($f_{\rm CC}=10,100$). In the asymmetric case, collision terms associated with $\Gamma^{\rm CC}-\bar{\Gamma}^{\rm CC}$ can couple the time evolution of $\mathbf{D}_{n}=~\average{(\mathbf{P}-\mathbf{\bar{P}})\cos^{n}\theta}$ and that of $\mathbf{S}_{m}=~\average{(\mathbf{P}+\mathbf{\bar{P}})\cos^{m}\theta}$ to each other \cite{Johns:2021arXiv210411369J}. Such coupling may make a qualitative difference between the collision effect in the symmetric case and that of the asymmetric case. The asymmetry of the collision rate is quite possible in explosive astrophysical sites. Here we focus on monochromatic energy neutrinos, but the asymmetry of the collision rate can be increased by considering the energy dependence of neutrinos.

%Such property of the asymmetry is consistent with the result in Ref.\cite{Johns:2021arXiv210411369J} irrespective of the absence of absorption and emission collision terms. Our results clarify the role of the asymmetry on the CC scatterings.

%Such similarity may come from the similarity between $C_{ee}$ in Eq.(\ref{eq:collision rho}) and $C^{\mathrm{CC}}_{ee}$ in Eq.(\ref{eq:collision rho johns}). These diagonal components may dominantly contribute to the enhancement of flavor conversions.

%%%%%%%%%%%%%%%%%%%%%%%%%%%%%%%%%%%%%%%%
\section{Conclusions}\label{sec:Conclusions}
%%%%%%%%%%%%%%%%%%%%%%%%%%%%%%%%%%%%%%%%

We calculate fast flavor conversions with collision effects of neutrino scatterings and analyze behaviors of the flavor conversions based on the dynamics of neutrino polarization vectors in cylindrical coordinate where we can easily access the information of the phase of rotation.  

We find that the the collision terms in Ref.~\cite{Shalgar:2020wcx} induce enhancement of flavor conversions and isotropization of neutrino distributions. In the linear phase, the instability of fast flavor conversions grows up around the ELN crossing where the phase difference $\delta$ takes an intermediate value in the range between $0$ and $-\pi$. In the limit cycle phase, the evolution track of neutrino polarization vector without collision effects is closed, and flavor conversions show a periodic trend. On the other hand, with the collision terms, the evolution track of the neutrino polarization vector is no longer closed because of the decrease of $|\delta|$ in every cycle.
The collision term breaks the symmetry between positive and negative $\delta$, and the induced imbalance enhances the total transition probability.
After the synchronization of neutrino polarization vectors, the value of $\delta$ finally converges to zero. At the end of the limit cycle, the evolution of $\average{P_{ex}}$ settles down to equilibrium, and the distributions of neutrinos are dependent on the neutrino scattering angle. Such properties are well consistent with the results in Ref.~\cite{Shalgar:2020wcx}. However, in the relaxation phase, all of the neutrino polarization vectors align with the $z$-axis keeping the value of  $\average{P_{ex}}$. The distributions of neutrinos finally become isotropic after the relaxation phase. 
%\textcolor{blue}{(Remove this sentence)Such isotropic neutrino distributions after the fast flavor conversions are also confirmed in Ref.~\cite{Martin:2019gxb,Martin:2021xyl}. }

Furthermore, we calculate the effect of collision terms used in previous studies, which helps unify presently disparate effects of collisional instability. For the NC scattering, the flavor conversions are enhanced in the small collision term. On the other hand, the large collision term prevents flavor conversions. Our results suggest that the collision effect is characterized by the ratio of the parameters between collision terms and neutrino-neutrino interactions. For the CC scattering, the transition probability is significantly raised up in a large asymmetry between neutrino and antineutrino collision rates. In the case of the asymmetric CC collision terms, the flavor conversions are not suppressed even in large collision parameters.

Here, we remark on the uncertainties of our works. The calculation results of fast flavor conversions are very sensitive to the numerical setup. The flavor conversions without collisions are highly periodic in our calculation assuming the spatial homogeneity and ignoring the azimuthal angle dependence. In some case, even without the collision term, fast flavor conversions can decay and reach a stationary solution \cite{Xiong:2021dex}. It has been shown that the flavor conversions become more chaotic and non-periodic in spatial inhomogeneous system \cite{Martin:2019gxb,Zaizen2021a}. In addition, the periodic structure of flavor conversions is broken in the calculation with the azimuthal angle dependence \cite{Richers:2021nbx,Shalgar:2021arXiv210615622S}. The significant enhancement of flavor conversions due to the collision effect may be obscured, when the assumption in our calculation is relaxed. Our simplified numerical setup should be updated in order to study the collision effect precisely in more realistic environment. Here, we focus on behaviors of fast flavor conversions of two flavor neutrinos for simplicity, but three flavors of neutrinos are required to predict the reliable neutrino signal in explosive astrophysical sites.

\begin{acknowledgments}
We thank E. Kokubo, M. Delfan Azari and L. Johns for fruitful discussions and useful comments. This work was carried out under the
auspices of the National Nuclear Security Administration of the
U.S. Department of Energy at Los Alamos National Laboratory under
Contract No.~89233218CNA000001.
This study was supported in part by JSPS/MEXT KAKENHI Grant Numbers 
JP18H01212, % Yokoi Kinban B (Takiwaki Co-PI)
JP17H06364, %Shingakujutu GWGEN C01,  (Takiwaki Co-PI).
JP21H01088.
This work is also supported by the NINS program for cross-disciplinary
study (Grant Numbers 01321802 and 01311904) on Turbulence, Transport,
and Heating Dynamics in Laboratory and Solar/Astrophysical Plasmas:
"SoLaBo-X”, and also by MEXT as “Program for Promoting 
researches on the Supercomputer Fugaku” (Toward a unified view of 
the universe: from large scale structures to planets, JPMXP1020200109) with JICFuS
Numerical computations were carried out on PC cluster at the Center for Computational Astrophysics,
National Astronomical Observatory of Japan.
\end{acknowledgments}

\appendix
%%%%%%%%%%%%%%%%%%%%%%%%%%%%%%%%%%%%%%

\section{Synchronization in linear phase}\label{sec:slp 1}
%%%%%%%%%%%%%%%%%%%%%%%%%%%%%%%%%%%%%%
Here we focus on the phase $\delta$ seen in Fig.~\ref{fig:phase00905_caseb} in detail. The time derivative is written as
\begin{equation}
\frac{{\rm d} \delta}{{\rm d} t}
=H_z-\frac{{\rm d} H_\phi}{{\rm d} t} - \left(\frac{P_z}{P_R}\right)H_R\cos\delta
%=\Gamma(\delta)
,\label{eq:phase-eq 1}
 \end{equation}
where we ignore the collision term, which is not important in the linear phase. When the synchronization condition,
\begin{equation}
    \frac{{\rm d} \delta}{{\rm d} t}=0,\label{eq:synchronization1}
 \end{equation}
is satisfied, the profile of $\delta$ is derived from Eq.~(\ref{eq:phase-eq 1}),
\begin{equation}
    \delta \sim \cos^{-1}\left[
    \frac{H_z-\frac{\mathrm{d}{H}_\phi}{\mathrm{d}t}}{H_R} \left(\frac{P_R}{P_z}\right)
    \right].\label{eq:synchronization2}
 \end{equation}
The value of $\delta$ indicates the susceptibility to the flavor conversion. 
From the above equation, we obtain the following condition for $\delta=\pm\frac{\pi}{2}$,
 \begin{equation}
 \label{eq:condition of the significant conversions}
     H_z=\frac{\mathrm{d}{H}_\phi}{\mathrm{d}t}.
 \end{equation}
At this point, the significant flavor conversion is expected from Eq.~\eqref{eq:time evolution Pz in the linear phase}.

We confirmed that the synchronization condition of Eq.~(\ref{eq:synchronization1}) holds in the present calculation. In fact, Eq.~(\ref{eq:synchronization2}) reproduces well the result of $\delta$ in Fig.~\ref{fig:phase00905_caseb}. The value of $\delta$ is negative because the synchronization is stable, $\frac{\partial}{\partial \delta}(\frac{{\rm d} \delta}{{\rm d} t})< 0$. In addition, Eq.~(\ref{eq:condition of the significant conversions}) is satisfied at $\cos\theta\sim0.4$ in the linear phase. In terms of synchronizing phenomena, we can predict the angle $\theta$ at which flavor conversions will occur.

The analysis above is based on the textbook of synchronization phenomena (Section~4.10 of Ref.~\citep{Kuramoto2017}, which is written in Japanese),
many types of phase equation are solved in the book.
In the case of the coupling two oscillators,
\begin{align}
    \frac{{\rm d}\phi_1}{{\rm d}t} &= \omega_1 + \Gamma(\phi_1-\phi_2),\\
    \frac{{\rm d}\phi_2}{{\rm d}t} &= \omega_2 + \Gamma(\phi_2-\phi_1),
\end{align}
it is useful to take the phase difference $\psi=\phi_1-\phi_2$,
\begin{align}
    \frac{{\rm d}\psi}{{\rm d}t} &= \Gamma_a(\psi),
\end{align}
where $\Gamma_a$ contains the asymmetric part of $\Gamma$ and the contribution of $\omega_1-\omega_2$. 
The condition for the synchronization is given by $\Gamma_a=0$ and $\frac{{\partial}\Gamma_a}{\partial \psi}<0$.
The second condition ensures the stability for small perturbation on $\psi$, i.e.,
$\frac{{\rm d}\psi}{{\rm d}t}$ is negative (positive) for positive (negative) perturbations.
This general argument is also applicable to our system as $\phi_1=P_\phi$, $\phi_2 = H_\phi$ and $\psi=\delta$.

%%%%%%%%%%%%%%%%%%%%%%%%%%%%%%%%%%%%%%
\section{The derivation of Eq.~(\ref{eq:Time evolution of delta})}\label{sec:derivation of time evolution of delta}
%%%%%%%%%%%%%%%%%%%%%%%%%%%%%%%%%%%%%%
%Here, we derive Eq.~\eqref{eq:Time evolution of delta}. 
As discussed in Sec.~\ref{sec:Relaxation phase}, both $\mathbf{D_{0}}$ and $\mathbf{B}$ are almost parallel to the $z$-axis when vacuum frequency and the mixing angle are small. In such a case, the polarization vector of neutrino Hamiltonian on the $x$-$y$ plane is described by
\begin{align}
\label{eq:Hamiltonian vector x}
    H_{x}&\sim-\mu^{\prime}\cos\theta D_{1x},
\end{align}
\begin{align}
\label{eq:Hamiltonian vector y}
    H_{y}&\sim-\mu^{\prime}\cos\theta D_{1y},
\end{align}
where $D_{1x}$ and $D_{1y}$ are $x$ and $y$ components of $\mathbf{D_{1}}$ in Eq.~\eqref{eq:vectors}. Then, the equation of motion of $H_{i}(i=x,y)$ is written as
\begin{equation}
\label{eq:time evolution of Hi}
    \frac{\mathrm{d}H_{i}}{\mathrm{d}t}\sim
    \left.\frac{\mathrm{d}H_{i}}{\mathrm{d}t}\right|_{C=0}-C H_{i},
\end{equation}
where the first term on the right hand side of Eq.~\eqref{eq:time evolution of Hi} does not include the collision parameter $C$ explicitly. From Eq.~\eqref{eq:time evolution of Hi}, the time derivative of the phase ${H}_{\phi}$ is given by
\begin{equation}
\label{eq:time evolution of Hphi}
\frac{\mathrm{d}H_{\phi}}{\mathrm{d}t}=-\frac{1}{H_{R}^{2}}\left(H_{y}\frac{\mathrm{d}H_{x}}{\mathrm{d}t} - H_{x}\frac{\mathrm{d}H_{y}}{\mathrm{d}t}\right)\sim\left.\frac{\mathrm{d}{H}_{\phi}}{\mathrm{d}t}\right|_{C=0}.
\end{equation}
Finally, Eq.~\eqref{eq:Time evolution of delta} is derived from Eqs.~\eqref{eq:time evolution Pp in the limit cycle phase} and \eqref{eq:time evolution of Hphi}.

\bibliography{ref,ref2}


\end{document}